\documentclass[aps,prd,onecolumn,preprintnumbers,notitlepage,superscriptaddress,
nofootinbib,amsfont,amssymb,10pt,singlespacing] {revtex4-1}
\usepackage{amsmath,bm}
\usepackage{times}
\usepackage{braket}
\usepackage{color,graphicx}
\usepackage{slashed}
\usepackage{hyperref}
\usepackage{graphics}
\usepackage{subfigure}
\usepackage{multirow,makecell}
\usepackage{textcomp}
\usepackage{mathtools}
\usepackage{float}

\newcommand{\beq}{\begin{eqnarray}}
\newcommand{\eeq}{\end{eqnarray}}
\newcommand{\non}{\nonumber\\ }

\def\lsim{ {\ \lower-1.2pt\vbox{\hbox{\rlap{$<$}\lower6pt\vbox{\hbox{$\sim$}
}}}\ } }
\def\gsim{ {\ \lower-1.2pt\vbox{\hbox{\rlap{$>$}\lower6pt\vbox{\hbox{$\sim$}
}}}\ } }
\definecolor{Red}{rgb}{1.,0.,0.}

\definecolor{Blue}{rgb}{0.,0.,1.}

\definecolor{nicered}{rgb}{0.7,0.1,0.1}
\definecolor{nicegreen}{rgb}{0.1,0.5,0.1}

\hypersetup{colorlinks,citecolor=nicegreen,linkcolor=nicered}

\begin{document}

\title{
The perturbative QCD predictions for the decay ${B^{0}_{s}}\rightarrow SS(a_{0}(980),f_{0}(980),f_{0}(500))$}
%%%==================================================================

\author{Ze-Rui~Liang}
\email[Electronic address:]{liangzr@email.swu.edu.cn}
\affiliation{School of Physical Science and Technology,
Southwest University, Chongqing 400715, China}

\author{Xian-Qiao~Yu}
\email[Electronic address:]{yuxq@swu.edu.cn}
\affiliation{School of Physical Science and Technology,
Southwest University, Chongqing 400715, China}

\date{\today}

%%%%%%%%%%%%%%%%%%%%%%%%%%%%%%%%%%%%%%%%%%%%%%%%%%%%%%%%%%%%%%%%%%
\begin{abstract}
In this work, we calculate the branching ratios and CP violations of the ${B^{0}_{s}}\rightarrow a_{0}(980)a_{0}(980)$ decay modes with both charged and neutral $a_{0}(980)$ mesons and ${B^{0}_{s}}\rightarrow f_{0}(980)(f_{0}(500))f_{0}(980)(f_{0}(500))$ for the first time in the pQCD approach. Considering the recent observation of the BESIII collaboration that provide a direct information about the constituent two-quark components in the corresponding $a_{0}(980)$
wave functions, we regard the scalar mesons $a_{0}(980)$, $f_{0}(980)$ and $f_{0}(500)$ as the $q\bar{q}$ quark component in our present work, and then make predictions of these decay modes. The branching ratios of our calculations are at the order of the $10^{-4}\sim10^{-6}$ when we consider the mixing scheme. We also calculate the CP violation parameters of these decay modes. The relatively large branching ratios make it easily to be tested by the running LHC-b experiments, and it can help us to understand both the inner properties and the QCD behavior of the scalar meson.
\end{abstract}
%%%%%%%%%%%%%%%%%%%%%%%%%%%%%%%%%%%%%%%%%%%%%%%%%%%%%%%%%%%%%%

\maketitle

%
%%%
%%%%%%%%%%%%%%%%% I. INTRODUCTION %%%%%%%%%%%%%%%%%%%%%%%%%%%%%%%%
%%%
%

\section{Introduction}\label{sec:intro}

 Since the first scalar meson $f_{0}(980)$ was observed by the Belle collaboration in the charged decay mode $B^{\pm}\rightarrow K^{\pm}f_{0}(980)\rightarrow K^{\pm}\pi^{\mp}\pi^{\pm}$~\cite{Abe:2002av}, and afterwards confirmed by BaBar~\cite{Aubert:2003mi}, a lot of other scalar mesons have been discovered in the experiment successively. The scalar mesons, especially for the $a_{0}(980)$ and $f_{0}(980)$, which are important for understanding the chiral symmetry and confinement in the low-energy region, are one of the key problems in the nonperturbative QCD~\cite{Achasov:2017zhy}. However, the inner structure of scalar mesons is still a contradiction in both the theoretical and experimental side, and many works have been done about scalar mesons in order to solve this problem~\cite{Liu:2013cvx, Dou:2015mka, Lu:2006fr, Colangelo:2010bg, Wang:2006ria, Cheng:2005nb, Liu:2010kq, Zou:2017yxc, Cheng:2019tgh, Li:2019jlp, Liu:2013lka, Cheng:2007st, Alford:2000mm}. In Ref.~\cite{Achasov:2017zhy}, the authors listed many evidences that sustain the four-quark model of the light scalar mesons based on a series of experimental data. In Ref.~\cite{Shen:2006ms}, the predicted result of $B\rightarrow a_{0}(980)K$ was $2$ times difference from the experimental result, and the author conclude that $a_{0}(980)$ cannot be interpreted as $q\bar{q}$. In Ref.~\cite{Weinstein:1990gu}, the authors showed that the production of the $S^{*}$ and $\delta$ and of low-mass $K\bar{K}$ pairs have properties of the $K\bar{K}$ molecules. Moreover, the scalar meson are identified as the quark-antiquark gluon hybrid. Nevertheless, these interpretations of the scalar mesons make theoretical calculations difficult, apart from the ordinary $q\bar{q}$ model.

In theoretical side, there are two interpretations about light scalar mesons below $2$ GeV in Review of Particle Physics~\cite{Tanabashi:2018oca}, the scalars below $1$ GeV, including $f_{0}(500)$, $K^{*}(700)$, $f_{0}(980)$ and $a_{0}(980)$ , form a $SU(3)$ flavor nonet, and $f_{0}(1370)$, $a_{0}(1450)$, $K^{*}(1430)$ and $f_{0}(1500)$ (or $f_{0}(1700)$) that above $1$ GeV form another $SU(3)$ flavor nonet. In order to describe the structure of these light scalar mesons , the authors of Ref.~\cite{Cheng:2005nb} presented two Scenarios to clarify the scalar mesons:

(1) Scenario~1, the light scalar mesons, which involved in the first $SU(3)$ flavor nonet, are usually regarded as the lowest-lying $q\bar{q}$ states, and the other nonet as the relevant first excited states. In the ordinary diquark model, the quark components of $a_{0}(980)$ and $f_{0}(980,500)$ are
\begin{equation}
a^{+}_{0}(980)=u\bar{d}, a^{-}_{0}(980)=\bar{u}d, a^{0}_{0}(980)=\frac{1}{\sqrt{2}}(u\bar{u}-d\bar{d}),
f_{0}(980)=s\bar{s}, f_{0}(500)=\frac{1}{\sqrt{2}}(u\bar{u}+d\bar{d}),
\end{equation}

(2) Scenario~2, the scalar mesons in the second nonet are regarded as the ground states($q\bar{q}$), and scalar mesons with mass between $2.0\sim2.3$ GeV are first excited states. This Scenario indicate that the scalars below or near $1$ GeV are four-quark bound states, while other scalars consist of $q\bar{q}$ in Scenario~1. So the quark components of $a_{0}(980)$ and $f_{0}(980,500)$ are
\begin{equation}
a^{+}_{0}(980)=u\bar{d}s\bar{s},    a^{-}_{0}(980)=\bar{u}d\bar{s}s,      a^{0}_{0}(980)=\frac{1}{\sqrt{2}}(u\bar{u}-d\bar{d})s\bar{s},
f_{0}(980)=\frac{1}{\sqrt{2}}(u\bar{u}+d\bar{d})s\bar{s},        f_{0}(500)=ud\bar{u}\bar{d}.
\end{equation}

Recently, the BESIII collaboration declare that the first
measurement of D mesons semileptonic decay $D^{0} \rightarrow d\bar{u}e^{+}\nu \rightarrow a^{-}_{0}(980)e^{+}\nu \rightarrow \pi^{-}\eta e^{+}\nu$ and the existing evidence of $D^{+} \rightarrow d\bar{d}e^{+}\nu \rightarrow a^{0}_{0}(980)e^{+}\nu \rightarrow \pi^{0}\eta e^{+}\nu$~\cite{Ablikim:2018ffp}, which would provide useful information
on revealing the mysterious nature of the scalar mesons. And in Ref.~\cite{Ablikim:2018pik}, BES III declare the $a^{0}_{0}(980)$-$f_{0}(980)$ mixing in the $J/\psi\rightarrow\phi f_{0}(980)\rightarrow\phi a^{0}_{0}(980)\rightarrow \phi \eta \pi^{0}$ and $\chi_{c1} \rightarrow a^{0}_{0}(980)\pi^{0} \rightarrow f_{0}(980)\pi^{0} \rightarrow \pi^{+}\pi^{-}\pi^{0}$ decay modes, which is the first observation of $a^{0}_{0}(980)$-$f_{0}(980)$ mixing in experiment. In our work, we treat the scalar mesons $a_{0}(980)$, $f_{0}(980)$ as the component of $q\bar{q}$ in scenario~1, and make the theoretical calculations within the perturbative QCD approach. For $f_{0}(980)$, there exist a mixing with the $f_{0}(500)$ in the SU(3) nonet, and in this work, we also take the mixing effect into account to make more reliable results. Motivated by the uncertain inner structure of the scalar mesons and very few works about the $B\rightarrow SS$ decays ($S$ denote the scalar mesons) to be studied in these general factorization approaches, we explore the branching ratios and CP-violating asymmetries of decay modes $\bar{B}^{0}_{s}\rightarrow a_{0}(980)a_{0}(980)$ and  $\bar{B}^{0}_{s}\rightarrow f_{0}(980,500)f_{0}(980,500)$~\footnote{$a_{0}(980)$, $f_{0}(980)$ and $f_{0}(500)$ will be respectively abbreviated as $a_{0}$, $f_{0}$ and $\sigma$ in the last part.} in perturbative QCD approach within the traditional two-quark model for the first time. Because the LHC-b collaboration are collecting more and more B mesons decays data, so we believe that our results can be testified by the experiment in the near future time.

This article is organized roughly in this order: in Section \ref{sec:pert}, we give a theoretical framework of the pQCD, list the wave functions that we need in the calculations, and also the perturbative calculations; in Section \ref{sec:numer}, we make numerical calculations and some discussions for the results that we get; and at last, we summary our work in the final Section. Some formulae what we used in our calculation are collected in the Appendix.

%
%%%
%%%%%%%%%%%%%%%%% II. Theoretical frame and the wave function %%%%%%%%%%%%%%%%%%%%%%%%%%%%%%%%
%%%
%
\section{The Theoretical Framework And Perturbative Calculation}\label{sec:pert}

The pQCD approach have been widely applied to calculate the hadronic matrix elements in the B mesons decay modes, it is based on the $k_{T}$ factorization. The divergence of the end-point singularity can be safely avoided by preserving the transverse momenta $k_{T}$ in the valence quark, and the only input parameters are the wave functions of the involved mesons in this method. Then the transition form factors and the different contributions, whose may contain the spectator and annihilation diagrams, are all calculated in this framework.

\subsection{Wave Functions and Distribution Amplitudes}

In kinematics aspects, we adopt the light-cone coordinate system in our calculation. Assuming the $B^{0}_{s}$ meson to be rest in the system, we can describe the momenta of the mesons in light-cone coordinate system, where the momenta are expressed in the form of $(p^{+}, p^{-}, p_{T})$ with the definition $p^{\pm}=\frac{p_{0}\pm p_{3}}{\sqrt{2}}$ and $p_{T}=(p_{1}, p_{2})$.

In our calculation, the wave function of the hadron $B^{0}_{s}$ can be found in Refs.~\cite{Lu:2000em,Keum:2000wi,Keum:2000ph}
\begin{equation}
\Phi_{B^{0}_{s}}=\frac{i}{\sqrt{2{\emph{N}}_{c}}}(\not {p}_{B}+m_{B_{s}}){\gamma_{5}}{\phi_{B_{s}}({\emph{x}_{1},\emph{b}_{1}})},
\end{equation}
where the distribution amplitude(DA) ${\phi_{B_{s}}({\emph{x}_{1},\emph{b}_{1}})}$ of $B^{0}_{s}$ meson is written as mostly used form, which is
\begin{equation}
\phi_{B_{s}}({\emph{x}_{1},\emph{b}_{1}})=\emph{N}_{B}{{\emph{x}_{1}}^2}(1-{\emph{x}_{1}})^2\exp[-\frac{m^2_{B_{s}}{{\emph{x}_{1}}^2}}{2\omega^2_{B_{s}}}-\frac{1}{2}(\omega_{B_{s}}{\emph{b}_{1}})^2],
\end{equation}
the normalization factor $\emph{N}_{B}=62.8021$ can be calculated by the normalization relation $\int^{1}_{0}\emph{dx}\phi_{B_{s}}(\emph{x}_{1},\emph{b}_{1}=0)=\emph{f}_{B_{s}}/({2}{\sqrt{{2}{\emph{N}}_{c}}})$ with ${\emph{N}}_{c}=3$ is the color number and decay constant $f_{B_{s}}=227.2 \pm 3.4$ {\rm MeV}. Here, we choose shape parameter $\omega_{B_{s}}=0.50\pm0.05$ {\rm GeV}~\cite{Ali:2007ff}.

For the scalar meson $a_{0}(980)$ and $f_{0}(980)$, the wave function can be read as~\cite{Cheng:2005nb,Cheng:2007st}:
\begin{equation}
\Phi_{S}(\emph{x})=\frac{1}{2\sqrt{2 \emph{N}_{c}}}[\not {p}\phi_{S}(\emph{x})+m_{S}\phi^{S}_{S}(\emph{x})+m_{S}(\not {v}\not{n}-1)\phi^{T}_{S}(\emph{x})],
\end{equation}
where $x$ denotes the momentum fraction of the meson, and $n=(1,0,0_{T})$, $v=(0,1,0_{T})$ are light-like dimensionless vectors.

 $\phi_{S}$ is the leading-twist distribution amplitude, the explicit form of which is expanded by the Gegenbauer polynomials~\cite{Cheng:2005nb, Cheng:2007st}:
\begin{equation}
\phi_{S}(\emph{x},\mu)=\frac{3}{\sqrt{2 {\emph{N}}_{c}}}\emph{x} (1-\emph{x})\{f_{S}(\mu)+\bar{f}_{S}(\mu)\sum^{\infty}_{m=1,3}B_{m}(\mu)C^{3/2}_{m}(2\emph{x}-1)\},
\end{equation}
and for the twist-3 DAs $\phi^{S}_{S}$ and $\phi^{T}_{S}$, we adopt the asymptotic forms in our calculation,
\begin{gather}
\phi^{S}_{S}(\emph{x},\mu)=\frac{1}{2 \sqrt{2{\emph{N}}_{c}}}\bar{f}_{S}(\mu),\\
\phi^{T}_{S}(\emph{x},\mu)=\frac{1}{2 \sqrt{2{\emph{N}}_{c}}}\bar{f}_{S}(\mu)(1-2\emph{x}),
\end{gather}
where $f_{S}$ and $\bar{f}_{S}$ are the vector and scalar decay constants of the scalar mesons $a_{0}$ and $f_{0}$ respectively, $B_{m}$ is Gegenbauer moment and $C^{3/2}_{m}(2\emph{x}-1)$ in DA of $\phi_{S}$ is Gegenbauer polynomials, these parameters are scale-dependent. A lot of calculations have been carried out about the light scalar mesons in various model~\cite{Brito:2004tv, Maltman:1999jn, Shakin:2001sz}. In this article, we adopt the value for decay constants and Gegenbauer moments in the DAs of the $a_{0}$ and $f_{0}$ as listed follow, which were calculated in QCD sum rules at the scale $\mu=1$ {\rm GeV}~\cite{Cheng:2005nb, Cheng:2007st}:
\begin{equation}
\begin{split}
&\bar{f}_{a_{0}}=0.365 \pm 0.020 {\rm GeV},  B_{1}=-0.93\pm 0.10, B_{3}=0.14\pm 0.08;\\
&\bar{f}_{S}=\bar{f}_{f_0}^n=\bar{f}_{f_0}^s=0.370 \pm 0.020 {\rm GeV},  B^{n}_{1}=-0.78\pm 0.08, B^{n}_{3}=0.02\pm 0.07, \\
&B^{s}_{1,3}=0.8B^{n}_{1,3}.
\end{split}
\end{equation}
The two decay constants $\bar{f}_{f_0}^n$ and $\bar{f}_{f_0}^s$ used in our calculations have been defined in the framework of the QCD sum rule method, here we choose the same value of these two constants and the reasons have been discussed in the Ref.~\cite{Cheng:2005nb}.
It is noticeable that only the odd Gegenbauer moments are taken into account due to the conservation of vector current or charge conjugation invariance. And we also pay attention to only the Gegenbauer moments $B_{1}$ and $B_{3}$ because the higher order Gegenbauer moments make tiny contributions and can be ignored safely.

The vector and scalar decay constants satisfy the relationship
\begin{equation}
\bar{f}_{S}(\mu)=\mu_{S}{f}_{S}(\mu)
\end{equation}
with
\begin{equation}
\mu_{S}=\frac{m_S}{m_{1}(\mu)-m_{2}(\mu)},
\end{equation}
and $m_{S}$ is the mass of the scalar meson and $m_{1}$ and $m_{2}$ are the running current quark masses in the scalar meson. From the above relationship, it is clear to see that the vector decay constant is proportional to the mass difference between the $m_{1}$ and $m_{2}$ quark, the mass difference is so small after considering the $SU(3)$ symmetry breaking that would heavily suppress the vector decay constant, which lead to the vector decay constants of the scalar mesons are very small and can be negligible. Likewise, for the same reason that only the odd Gegenbauer momentums are considered, the neutral scalar mesons can not be produced by the vector current, so in this work we adopt the vector constant $f_{S}=0$.

And the normalization relationship of the twist-2 and twist-3 DAs are
\begin{equation}
\begin{split}
\int^{1}_{0}\emph{dx}\phi_{S}(\emph{x})&=\int^{1}_{0}\emph{dx}\phi^{T}_{S}(\emph{x})=0,\\
\int^{1}_{0}\emph{dx}\phi^{S}_{S}(\emph{x})&=\frac{\bar{f}_{S}}{2\sqrt{2{\emph{N}}_{c}}}.
\end{split}
\end{equation}

For the scalar meson $f_{0}$-$\sigma$ system, the mixing should have the relation:
 \beq
\left(
\begin{array}{c} \sigma \\ f_0 \\ \end{array} \right ) &=&
  \left( \begin{array}{cc}
 \cos{\theta} & -\sin{\theta} \\
 \sin{\theta} & \cos{\theta} \end{array} \right )
 \left( \begin{array}{c}  f_n\\ f_s \\ \end{array} \right )\;.
 \label{mix-f0}
 \eeq

\subsection{Perturbative Calculations}

For $\bar{B}^{0}_{s}\rightarrow SS$ decay mode, the relevant weak effective Hamiltonian can be written as~\cite{Buchalla:1995vs}

\begin{equation}
{\cal H}_{eff}=\frac{G_{F}}{\sqrt2}\big\{V_{ub}V^*_{us}[C_{1}(\mu)O_{1}(\mu)+C_{2}(\mu)O_{2}(\mu)]-V_{tb}V^*_{ts}[\sum^{10}_{i=3}C_{i}(\mu)O_{i}(\mu)]\big\},
\end{equation}
where $G_{F}=1.66378\times 10^{-5}$ $\mathrm{GeV^{-2}}$ is Fermi constant, and $V_{ub}$$V^*_{us}$ and $V_{tb}$$V^{*}_{ts}$ are Cabibbo-Kobayashi-Maskawa (CKM) factors, $O_{i}(\mu)$ ($i=1,2,...,10$) is local four-quark operator, which will be listed as follows, and $C_{i}(\mu)$ is corresponding Wilson coefficient.

(1) Current-Current Operators (Tree):
\begin{equation}
\begin{split}
O_{1}=(\bar{s}_{\alpha}u_{\beta})_{V-A}(\bar{u}_{\beta}b_{\alpha})_{V-A}, \quad O_{2}=(\bar{s}_{\alpha}u_{\alpha})_{V-A}(\bar{u}_{\beta}b_{\beta})_{V-A},
\end{split}
\end{equation}

(2) QCD Penguin Operators:
\begin{equation}
\begin{split}
O_{3}=(\bar{s}_{\alpha}b_{\alpha})_{V-A}\sum_{q}(\bar{q}_{\beta}q_{\beta})_{V-A},\quad O_{4}=(\bar{s}_{\alpha}b_{\beta})_{V-A}\sum_{q}(\bar{q}_{\beta}q_{\alpha})_{V-A},\\
O_{5}=(\bar{s}_{\alpha}b_{\alpha})_{V-A}\sum_{q}(\bar{q}_{\beta}q_{\beta})_{V+A},   \quad O_{6}=(\bar{s}_{\alpha}b_{\beta})_{V-A}\sum_{q}(\bar{q}_{\beta}q_{\alpha})_{V+A},
\end{split}
\end{equation}

(3) Electroweak Penguin Operators:
\begin{equation}
\begin{split}
O_{7}=\frac{3}{2}(\bar{s}_{\alpha}b_{\alpha})_{V-A}\sum_{q}e_{q}(\bar{q}_{\beta}q_{\beta})_{V+A},\quad O_{8}=\frac{3}{2}(\bar{s}_{\alpha}b_{\beta})_{V-A}\sum_{q}e_{q}(\bar{q}_{\beta}q_{\alpha})_{V+A},\\
O_{9}=\frac{3}{2}(\bar{s}_{\alpha}b_{\alpha})_{V-A}\sum_{q}e_{q}(\bar{q}_{\beta}q_{\beta})_{V-A}, \quad   O_{10}=\frac{3}{2}(\bar{s}_{\alpha}b_{\beta})_{V-A}\sum_{q}e_{q}(\bar{q}_{\beta}q_{\alpha})_{V-A},
\end{split}
\end{equation}
with the color indices $\alpha$, $\beta$ and $(q\bar{q})_{V\pm A}=\bar{q}\gamma_{\mu}(1\pm\gamma_{5})q$. The $q$ denotes the $u$ quark and $d$ quark, and $e_{q}$ is corresponding charge.

\begin{figure}[htbp]
 \begin{tabular}{l}
 \includegraphics[width=0.5\textwidth]{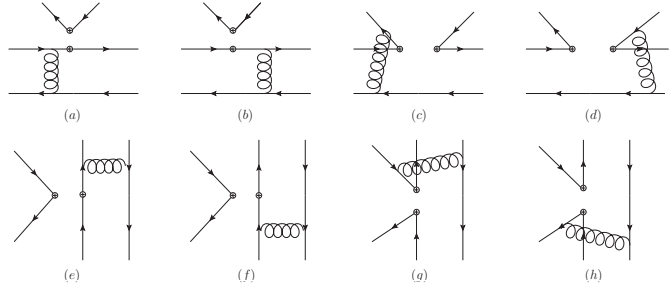}
 \end{tabular}
 \caption{The lowest order Feynman diagrams of the $\bar{B}^{0}_{s}\rightarrow SS$ decays in pQCD approach. The $\bar{B}^{0}_{s}\rightarrow a_{0}a_{0}$ decay is the rare decay mode, which only have the last line Feynman diagrams.}
   \label{fig:figure1}
 \end{figure}

The momenta of the $\bar{B}^{0}_{s}$, scalar mesons $M_{1}$, $M_{2}$ in the light-cone coordinate read as
\begin{equation}
\begin{split}
p_{B}=p_{1}=\frac{m_{B_{s}}}{\sqrt{2}}(1,1,0_{T}),                 \\
p_{2}=\frac{m_{B_{s}}}{\sqrt{2}}(r_{S}^2,1-r_{S}^2,0_{T}),              \\
p_{3}=\frac{m_{B_{s}}}{\sqrt{2}}(1-r_{S}^2,r_{S}^2,0_{T}),
\end{split}
\end{equation}
with the $B^{0}_{s}$ mass $m_{B_{s}}$ and the mass ratio $r_{S}=\frac{m_{S}}{m_{B_{s}}}$.

And the corresponding light quark's momenta in each meson read as
\begin{equation}
\begin{split}
k_{1}&=(\emph{x}_{1}p^{+}_{1},0,k_{1T})=(\frac{m_{B_{s}}}{\sqrt{2}}\emph{x}_{1},0,k_{1T}),\\
k_{2}&=(0,\emph{x}_{2}p^{-}_{2},k_{2T})=(0,\frac{m_{B_{s}}}{\sqrt{2}}(1-r_{S}^2)\emph{x}_{2},k_{2T}),\\
k_{3}&=(\emph{x}_{3}p^{+}_{3},0,k_{3T})=(\frac{m_{B_{s}}}{\sqrt{2}}(1-r_{S}^2)\emph{x}_{3},0,k_{3T}).
\end{split}
\end{equation}

Then based on the pQCD approach, we can write the decay amplitude as
\begin{equation}
{\cal A}\sim\int\emph{d}\emph{x}_{1}\emph{d}\emph{x}_{2}\emph{d}\emph{x}_{3}\emph{b}_{1}\emph{d}\emph{b}_{1}\emph{b}_{2}\emph{d}\emph{b}_{2}\emph{b}_{3}\emph{d}\emph{b}_{3}\times \mathrm{Tr}[ H(\emph{x}_{i}, \emph{b}_{i}, t) C_{t}
\Phi_{B}(\emph{x}_{1}, \emph{b}_{1})\Phi_{S}(\emph{x}_{2}, \emph{b}_{2}) \Phi_{S}(\emph{x}_{3}, \emph{b}_{3})S_{t}(\emph{x}_{i})e^{-S(t)}],
\end{equation}
where $\emph{b}_{i}$ is the conjugate momenta of $k_{i}$, and $t$ is the largest energy scale in hard function $H(\emph{x}_{i}, \emph{b}_{i}, t)$. The $e^{-S(t)}$ suppress the soft dynamics~\cite{Li:1997un} and make a reliable perturbative calculation of the hard function $H$, which come from higher order radiative corrections to wave functions and hard amplitudes. $\Phi_{M}$ represent universal and channel independent wave function, which describes the hadronization of mesons.

As depicted in Fig.~\ref{fig:figure1}, we calculate all the contributed diagrams respectively. We use $F$ and $M$ denote the factorizable and non-factorizable contributions respectively, and the subscript $a$, $c$, $e$, $g$ denote the contributions of the Feynman diagrams (a) and (b), (c) and (d), (e) and (f), (g) and (h) and the superscript $LL$, $LR$, $SP$ is the $(V-A)(V-A)$, $(V-A)(V+A)$ and $(S-P)(S+P)$ vertex, respectively. The vertex $(S-P)(S+P)$ is the Fierz transformation of the $(V-A)(V+A)$.

First, the total contribution of the factorization diagrams (a) and (b) with different currents are

(1) $(V-A)(V-A)$
\begin{equation}
\begin{split}
{\emph F}^{LL}_{a}&=8\pi\emph{C}_{\emph{F}}f_{S}m^{4}_{B_{s}}\int^{1}_{0}\emph{d}\emph{x}_{1}\emph{d}\emph{x}_{2}\int^\infty_{0}\emph{b}_{1}\emph{b}_{2}\emph{d}\emph{b}_{1}\emph{d}\emph{b}_{2}\phi_{B}(\emph{x}_{1},\emph{b}_{1})\\
& \times \{[-(2+\emph{x}_{2})\phi_{S}(\emph{x}_{2})+r_{S}(1+2\emph{x}_{2})(\phi^{S}_{S}(\emph{x}_{2})+\phi^{T}_{S}(\emph{x}_{2}))]\\
& \times h^{1}_{a}(\emph{x}_{1}, \emph{x}_{2}, \emph{b}_{1}, \emph{b}_{2})E_{ef}(t^{1}_{a})S_{t}(\emph{x}_{2})+[2r_{S}\phi^{S}_{S}(\emph{x}_{2})]\\
& \times h^{2}_{a}(\emph{x}_{1}, \emph{x}_{2}, \emph{b}_{1}, \emph{b}_{2})E_{ef}(t^{2}_{a})S_{t}(\emph{x}_{1})\},
\end{split}
\end{equation}

(2) $(V-A)(V+A)$
\begin{equation}
\begin{split}
{\emph F}^{LR}_{a}={\emph F}^{LL}_{a},
\end{split}
\end{equation}

(3) $(S-P)(S+P)$
\begin{equation}
\begin{split}
{\emph F}^{SP}_{a}&=-16\pi\emph{C}_{\emph{F}}\bar{f}_{S}m^{4}_{B_{s}}r_{S}\int^{1}_{0}\emph{d}\emph{x}_{1}\emph{d}\emph{x}_{2}\int^\infty_{0}\emph{b}_{1}\emph{b}_{2}\emph{d}\emph{b}_{1}\emph{d}\emph{b}_{2}\phi_{B}(\emph{x}_{1},\emph{b}_{1})\\
& \times \{[r_{S}(5+\emph{x}_{2})\phi^{S}_{S}(\emph{x}_{2})+r_{S}(1-\emph{x}_{2})\phi^{T}_{S}(\emph{x}_{2})-3\phi_{S}(\emph{x}_{2})]\\
& \times h^{1}_{a}(\emph{x}_{1}, \emph{x}_{2}, \emph{b}_{1}, \emph{b}_{2})E_{ef}(t^{1}_{a})S_{t}(\emph{x}_{2})\\
&-[2r_{S}(1-x_{1})\phi^{S}_{S}(\emph{x}_{2})+x_{1}\phi_{S}(\emph{x}_{2})]\\
& \times h^{2}_{a}(\emph{x}_{1}, \emph{x}_{2}, \emph{b}_{1}, \emph{b}_{2})E_{ef}(t^{2}_{a})S_{t}(\emph{x}_{1})\},
\end{split}
\end{equation}
with the color factor $\emph{C}_{\emph{F}}=\frac{\emph{N}_{c}^2-1}{2\emph{N}_{c}}=\frac{4}{3}$. The factorization contribution of the $(V-A)(V-A)$ and $(V-A)(V+A)$ current are negelected because the vector decay constant is a small value and we take it as zero.

For non-factorization diagrams, the total contribution from (c) and (d) is:

(1) $(V-A)(V-A)$
\begin{equation}
\begin{split}
{\emph M}^{LL}_{c}&=\frac{32\pi\emph{C}_{\emph{F}}m^{4}_{B_{s}}}{\sqrt{2{\emph{N}}_{c}}}\int^{1}_{0}\emph{d}\emph{x}_{1}\emph{d}\emph{x}_{2}\emph{d}\emph{x}_{3}\int^\infty_{0}\emph{b}_{2}\emph{b}_{3}\emph{d}\emph{b}_{2}\emph{d}\emph{b}_{3}\phi_{B}(\emph{x}_{1},\emph{b}_{3})\\
& \times \{[(\emph{x}_{3}+\emph{x}_{1}-1)\phi_{S}(\emph{x}_{3})\phi_{S}(\emph{x}_{2})
+r_{S}(1-\emph{x}_{2})\phi_{S}(\emph{x}_{3})(\phi^{S}_{S}(\emph{x}_{2})-\phi^{T}_{S}(\emph{x}_{2}))]\\
& \times h^{1}_{c}(\emph{x}_{1}, \emph{x}_{2}, \emph{x}_{3}, \emph{b}_{2}, \emph{b}_{3})E_{nef}(t^{1}_{c})\\
&-[(\emph{x}_{1}+\emph{x}_{2}-\emph{x}_{3}-1)\phi_{S}(\emph{x}_{3})\phi_{S}(\emph{x}_{2})+r_{S}(1-\emph{x}_{2})\phi_{S}(\emph{x}_{3})(\phi^{S}_{S}(\emph{x}_{2})+\phi^{T}_{S}(\emph{x}_{2}))]\\
& \times h^{2}_{c}(\emph{x}_{1}, \emph{x}_{2}, \emph{x}_{3}, \emph{b}_{2}, \emph{b}_{3})E_{nef}(t^{2}_{c})\},
\end{split}
\end{equation}

(2) $(V-A)(V+A)$
\begin{equation}
\begin{split}
{\emph M}^{LR}_{c}&=\frac{32\pi\emph{C}_{\emph{F}}m^{4}_{B_{s}}}{\sqrt{2{\emph{N}}_{c}}}\int^{1}_{0}\emph{d}\emph{x}_{1}\emph{d}\emph{x}_{2}\emph{d}\emph{x}_{3}\int^\infty_{0}\emph{b}_{2}\emph{b}_{3}\emph{d}\emph{b}_{2}\emph{d}\emph{b}_{3}\phi_{B}(\emph{x}_{1},\emph{b}_{3})\\
& \times \{[r^2_{S}(2-\emph{x}_{1}-\emph{x}_{2}-\emph{x}_{3})\phi^{S}_{S}(\emph{x}_{3})\phi^{S}_{S}(\emph{x}_{2})
+r^2_{S}(\emph{x}_{1}-\emph{x}_{2}+\emph{x}_{3})\phi^S_{S}(\emph{x}_{3})\phi^{T}_{S}(\emph{x}_{2})\\
&+r^2_{S}(\emph{x}_{1}-\emph{x}_{2}+\emph{x}_{3})\phi^T_{S}(\emph{x}_{3})\phi^{S}_{S}(\emph{x}_{2})
+r^2_{S}(2-\emph{x}_{1}-\emph{x}_{2}-\emph{x}_{3})\phi^T_{S}(\emph{x}_{3})\phi^T_{S}(\emph{x}_{2})\\
&+r_{S}(1-x_{1}-x_{3})(\phi^{S}_{S}(\emph{x}_{3})-\phi^{T}_{S}(\emph{x}_{3}))\phi_{S}(\emph{x}_{2})]
 \times h^{1}_{c}(\emph{x}_{1}, \emph{x}_{2}, \emph{x}_{3}, \emph{b}_{2}, \emph{b}_{3})E_{nef}(t^{1}_{c})\\
&-[r^2_{S}(1-\emph{x}_{1}-\emph{x}_{2}+\emph{x}_{3})\phi^{S}_{S}(\emph{x}_{3})\phi^{S}_{S}(\emph{x}_{2})
+r^2_{S}(\emph{x}_{1}-\emph{x}_{2}-\emph{x}_{3}+1)\phi^S_{S}(\emph{x}_{3})\phi^{T}_{S}(\emph{x}_{2})\\
&-r^2_{S}(\emph{x}_{1}-\emph{x}_{2}-\emph{x}_{3}+1)\phi^T_{S}(\emph{x}_{3})\phi^{S}_{S}(\emph{x}_{2})
-r^2_{S}(1-\emph{x}_{1}-\emph{x}_{2}+\emph{x}_{3})\phi^T_{S}(\emph{x}_{3})\phi^T_{S}(\emph{x}_{2})\\
&+r_{S}(-x_{1}+x_{3})(\phi^{S}_{S}(\emph{x}_{3})+\phi^{T}_{S}(\emph{x}_{3}))\phi_{S}(\emph{x}_{2})]
 \times h^{2}_{c}(\emph{x}_{1}, \emph{x}_{2}, \emph{x}_{3}, \emph{b}_{2}, \emph{b}_{3})E_{nef}(t^{2}_{c})\},
\end{split}
\end{equation}

(3) $(S-P)(S+P)$
\begin{equation}
\begin{split}
{\emph M}^{SP}_{c}&=-\frac{32\pi\emph{C}_{\emph{F}}m^{4}_{B_{s}}}{\sqrt{2{\emph{N}}_{c}}}\int^{1}_{0}\emph{d}\emph{x}_{1}\emph{d}\emph{x}_{2}\emph{d}\emph{x}_{3}\int^\infty_{0}\emph{b}_{2}\emph{b}_{3}\emph{d}\emph{b}_{2}\emph{d}\emph{b}_{3}\phi_{B}(\emph{x}_{1},\emph{b}_{3})\\
& \times \{[(\emph{x}_{1}+\emph{x}_{2}+\emph{x}_{3}-2)\phi_{S}(\emph{x}_{3})\phi_{S}(\emph{x}_{2})
+r_{S}(1-\emph{x}_{2})\phi_{S}(\emph{x}_{3})(\phi^{S}_{S}(\emph{x}_{2})+\phi^{T}_{S}(\emph{x}_{2}))]\\
&\times h^{1}_{c}(\emph{x}_{1}, \emph{x}_{2}, \emph{x}_{3}, \emph{b}_{2}, \emph{b}_{3})E_{nef}(t^{1}_{c})\\
&-[(\emph{x}_{1}-\emph{x}_{3})\phi_{S}(\emph{x}_{3})\phi_{S}(\emph{x}_{2})
+r_{S}(1-\emph{x}_{2})\phi_{S}(\emph{x}_{3})(\phi^{S}_{S}(\emph{x}_{2})-\phi^{T}_{S}(\emph{x}_{2}))]\\
& \times h^{2}_{c}(\emph{x}_{1}, \emph{x}_{2}, \emph{x}_{3}, \emph{b}_{2}, \emph{b}_{3})E_{nef}(t^{2}_{c})\},
\end{split}
\end{equation}

The total contribution of the annihilation Feynman diagrams Fig.~\ref{fig:figure1} (e) and (f), which only involve the wave function of the final light scalar mesons, are

(1) $(V-A)(V-A)$
\begin{equation}
\begin{split}
{\emph F}^{LL}_{e}&=8\pi\emph{C}_{\emph{F}}f_{B}m^{4}_{B_{s}}\int^{1}_{0}\emph{d}\emph{x}_{2}\emph{d}\emph{x}_{3}\int^\infty_{0}\emph{b}_{2}\emph{b}_{3}\emph{d}\emph{b}_{2}\emph{d}\emph{b}_{3}\\
& \times \{[-\emph{x}_{3}\phi_{S}(\emph{x}_{3})\phi_{S}(\emph{x}_{2})+2r_{S}^2(1+\emph{x}_{3})\phi^{S}_{S}(\emph{x}_{3})\phi^{S}_{S}(\emph{x}_{2})-2r_{S}^2(1-\emph{x}_{3})\phi^{T}_{S}(\emph{x}_{3})\phi^{S}_{S}(\emph{x}_{2})]\\
& \times h^1_{e}(\emph{x}_{2}, \emph{x}_{3}, \emph{b}_{2}, \emph{b}_{3})E_{af}(t^1_{e})S_{t}(\emph{x}_{3})\\
&+[\emph{x}_{2}\phi_{S}(\emph{x}_{3})\phi_{S}(\emph{x}_{2})-2r_{S}^2(1+\emph{x}_{2})\phi^{S}_{S}(\emph{x}_{3})\phi^{S}_{S}(\emph{x}_{2})+2r_{S}^2(1-\emph{x}_{2})\phi^{S}_{S}(\emph{x}_{3})\phi^{T}_{S}(\emph{x}_{2})]\\
& \times h^2_{e}(\emph{x}_{2}, \emph{x}_{3}, \emph{b}_{2}, \emph{b}_{3})E_{af}(t^2_{e})S_{t}(\emph{x}_{2})\},
\end{split}
\end{equation}

(2) $(V-A)(V+A)$
\begin{equation}
\begin{split}
{\emph F}^{LR}_{e}={\emph F}^{LL}_{e},
\end{split}
\end{equation}

(3) $(S-P)(S+P)$
\begin{equation}
\begin{split}
{\emph F}^{SP}_{e}&=16\pi\emph{C}_{\emph{F}}f_{B}m^{4}_{B_{s}}\int^{1}_{0}\emph{d}\emph{x}_{2}\emph{d}\emph{x}_{3}\int^\infty_{0}\emph{b}_{2}\emph{b}_{3}\emph{d}\emph{b}_{2}\emph{d}\emph{b}_{3}\\
& \times \{[2r_{S}\phi_{S}(\emph{x}_{3})\phi^S_{S}(\emph{x}_{2})-r_{S}\emph{x}_{3}(\phi^{S}_{S}(\emph{x}_{3})-\phi^{T}_{S}(\emph{x}_{3}))\phi_{S}(\emph{x}_{2})]\\
& \times h^1_{e}(\emph{x}_{2}, \emph{x}_{3}, \emph{b}_{2}, \emph{b}_{3})E_{af}(t^1_{e})S_{t}(\emph{x}_{3})\\
&+[r_{S}\emph{x}_{2}\phi_{S}(\emph{x}_{3})(\phi^{S}_{S}(\emph{x}_{2})-\phi^{T}_{S}(\emph{x}_{2}))-2r_{S}\phi^{S}_{S}(\emph{x}_{3})\phi_{S}(\emph{x}_{2})]\\
& \times h^2_{e}(\emph{x}_{2}, \emph{x}_{3}, \emph{b}_{2}, \emph{b}_{3})E_{af}(t^2_{e})S_{t}(\emph{x}_{2})\},
\end{split}
\end{equation}

Then the total non-factorizable annihilation decay amplitudes for the Fig.~\ref{fig:figure1} (g) and (h) diagrams are

(1) $(V-A)(V-A)$
\begin{equation}
\begin{split}
{\emph M}^{LL}_{g}&=\frac{32\pi\emph{C}_{\emph{F}}m^{4}_{B_{s}}}{\sqrt{2{\emph{N}}_{c}}}\int^{1}_{0}\emph{d}\emph{x}_{1}\emph{d}\emph{x}_{2}\emph{d}\emph{x}_{3}\int^\infty_{0}\emph{b}_{1}\emph{b}_{2}\emph{d}\emph{b}_{1}\emph{d}\emph{b}_{2}\phi_{B}(\emph{x}_{1},\emph{b}_{1})\\
& \times \{[-\emph{x}_{2}\phi_{S}(\emph{x}_{2})\phi_{S}(\emph{x}_{3})-r_{S}^2(\emph{x}_{1}-\emph{x}_{3}-\emph{x}_{2})\phi^{S}_{S}(\emph{x}_{2})\phi^{S}_{S}(\emph{x}_{3})\\
& +r_{S}^2(\emph{x}_{1}-\emph{x}_{3}+\emph{x}_{2})\phi^{S}_{S}(\emph{x}_{2})\phi^{T}_{S}(\emph{x}_{3})+r_{S}^2(\emph{x}_{1}-\emph{x}_{3}+\emph{x}_{2})\phi^{T}_{S}(\emph{x}_{2})\phi^{S}_{S}(\emph{x}_{3})\\
&-r_{S}^2(\emph{x}_{1}-\emph{x}_{3}-\emph{x}_{2})\phi^{T}_{S}(\emph{x}_{2})\phi^{T}_{S}(\emph{x}_{3})]\\
& \times h^1_{g}(\emph{x}_{1}, \emph{x}_{2}, \emph{x}_{3}, \emph{b}_{1}, \emph{b}_{2})E_{naf}(t^1_{g})\\
&+[(\emph{x}_{1}+\emph{x}_{3})\phi_{S}(\emph{x}_{2})\phi_{S}(\emph{x}_{3})-r_{S}^2(2+\emph{x}_{1}+\emph{x}_{3}+\emph{x}_{2})\phi^{S}_{S}(\emph{x}_{2})\phi^{S}_{S}(\emph{x}_{3})\\
&+r_{S}^2(\emph{x}_{2}-\emph{x}_{1}-\emph{x}_{3})\phi^{S}_{S}(\emph{x}_{2})\phi^{T}_{S}(\emph{x}_{3})\\
&+r_{S}^2(\emph{x}_{2}-\emph{x}_{1}-\emph{x}_{3})\phi^{T}_{S}(\emph{x}_{2})\phi^{S}_{S}(\emph{x}_{3})+r_{S}^2(2-\emph{x}_{2}-\emph{x}_{1}-\emph{x}_{3})\phi^{T}_{S}(\emph{x}_{2})\phi^{T}_{S}(\emph{x}_{3})]\\
& \times h^2_{g}(\emph{x}_{1}, \emph{x}_{2}, \emph{x}_{3}, \emph{b}_{1}, \emph{b}_{2})E_{naf}(t^2_{g})\},
\end{split}
\end{equation}

(2) $(V-A)(V+A)$
\begin{equation}
\begin{split}
{\emph M}^{LR}_{g}&=\frac{32\pi\emph{C}_{\emph{F}}m^{4}_{B_{s}}}{\sqrt{2{\emph{N}}_{c}}}\int^{1}_{0}\emph{d}\emph{x}_{1}\emph{d}\emph{x}_{2}\emph{d}\emph{x}_{3}\int^\infty_{0}\emph{b}_{1}\emph{b}_{2}\emph{d}\emph{b}_{1}\emph{d}\emph{b}_{2}\phi_{B}(\emph{x}_{1},\emph{b}_{1})\\
& \times \{[r_{S}(\emph{x}_{1}-\emph{x}_{3})\phi_{S}(\emph{x}_{2})(\phi^S_{S}(\emph{x}_{3})+\phi^{T}_{S}(\emph{x}_{3})) -r_{S}\emph{x}_{2}(\phi^{S}_{S}(\emph{x}_{2})+\phi^{T}_{S}(\emph{x}_{2}))\phi_{S}(\emph{x}_{3})]\\
& \times h^1_{g}(\emph{x}_{1}, \emph{x}_{2}, \emph{x}_{3}, \emph{b}_{1}, \emph{b}_{2})E_{naf}(t^1_{g})\\
&+[r_{S}(\emph{x}_{1}+\emph{x}_{3}-2)\phi_{S}(\emph{x}_{2})(\phi^S_{S}(\emph{x}_{3})+\phi^{T}_{S}(\emph{x}_{3}))
-r_{S}(2-\emph{x}_{2})(\phi^{S}_{S}(\emph{x}_{2})+\phi^{T}_{S}(\emph{x}_{2}))\phi_{S}(\emph{x}_{3})]\\
& \times h^2_{g}(\emph{x}_{1}, \emph{x}_{2}, \emph{x}_{3}, \emph{b}_{1}, \emph{b}_{2})E_{naf}(t^2_{g})\},
\end{split}
\end{equation}

(3) $(S-P)(S+P)$
\begin{equation}
\begin{split}
{\emph M}^{SP}_{g}&=\frac{-32\pi\emph{C}_{\emph{F}}m^{4}_{B_{s}}}{\sqrt{2{\emph{N}}_{c}}}\int^{1}_{0}\emph{d}\emph{x}_{1}\emph{d}\emph{x}_{2}\emph{d}\emph{x}_{3}\int^\infty_{0}\emph{b}_{1}\emph{b}_{2}\emph{d}\emph{b}_{1}\emph{d}\emph{b}_{2}\phi_{B}(\emph{x}_{1},\emph{b}_{1})\\
& \times \{[(-\emph{x}_{1}+\emph{x}_{3})\phi_{S}(\emph{x}_{2})\phi_{S}(\emph{x}_{3})+r_{S}^2(\emph{x}_{1}-\emph{x}_{3}-\emph{x}_{2})\phi^{S}_{S}(\emph{x}_{2})\phi^{S}_{S}(\emph{x}_{3})\\
& +r_{S}^2(\emph{x}_{1}+\emph{x}_{2}-\emph{x}_{3})\phi^{S}_{S}(\emph{x}_{2})\phi^{T}_{S}(\emph{x}_{3})+r_{S}^2(\emph{x}_{1}+\emph{x}_{2}-\emph{x}_{3})\phi^{T}_{S}(\emph{x}_{2})\phi^{S}_{S}(\emph{x}_{3})\\
& +r_{S}^2(\emph{x}_{1}-\emph{x}_{3}-\emph{x}_{2})\phi^{T}_{S}(\emph{x}_{2})\phi^{T}_{S}(\emph{x}_{3})]\\
& \times h^1_{g}(\emph{x}_{1}, \emph{x}_{2}, \emph{x}_{3}, \emph{b}_{1}, \emph{b}_{2})E_{naf}(t^1_{g})\\
&+[-\emph{x}_{2}\phi_{S}(\emph{x}_{2})\phi_{S}(\emph{x}_{3})+r_{S}^2(2+\emph{x}_{1}+\emph{x}_{3}+\emph{x}_{2})\phi^{S}_{S}(\emph{x}_{2})\phi^{S}_{S}(\emph{x}_{3})\\
&-r_{S}^2(\emph{x}_{1}+\emph{x}_{3}-\emph{x}_{2})\phi^{S}_{S}(\emph{x}_{2})\phi^{T}_{S}(\emph{x}_{3})-r_{S}^2(\emph{x}_{1}+\emph{x}_{3}-\emph{x}_{2})\phi^{T}_{S}(\emph{x}_{2})\phi^{S}_{S}(\emph{x}_{3})\\
&+r_{S}^2(-2+\emph{x}_{1}+\emph{x}_{2}+\emph{x}_{3})\phi^{T}_{S}(\emph{x}_{2})\phi^{T}_{S}(\emph{x}_{3})]\\
&\times h^2_{g}(\emph{x}_{1}, \emph{x}_{2}, \emph{x}_{3}, \emph{b}_{1}, \emph{b}_{2})E_{naf}(t^2_{g})\},
\end{split}
\end{equation}
For the $\bar{B}^{0}_{s}\rightarrow a^{+}_{0}a^{-}_{0}$ decay, which is a rare decay mode and only have annihilation Feynman diagrams, the decay amplitude
of $\bar{B}^{0}_{s}\rightarrow a^{+}_{0}a^{-}_{0}$ decay is then
\begin{equation}
\begin{split}
{\cal{A}}(\bar{B}^{0}_{s}\rightarrow a^{+}_{0}a^{-}_{0})=V_{ub}V^*_{us}[C_{2}{\emph M}_{g}^{LL}]
 -V_{tb}V^{*}_{ts}[(2C_{4}+\frac{1}{2}C_{10}){\emph M}_{g}^{LL}+(2C_{6}+\frac{1}{2}C_{8}){\emph M}^{SP}_{g}]
\end{split}
 \label{amp}
\end{equation}

Meanwhile, the relationship with respect to the decay $\bar{B}^{0}_{s}\rightarrow a^{0}_{0}a^{0}_{0}$ is
\begin{equation}
\sqrt{2}{\cal{A}}(\bar{B}^{0}_{s}\rightarrow a^{0}_{0}a^{0}_{0})={\cal{A}}(\bar{B}^{0}_{s}\rightarrow a^{+}_{0}a^{-}_{0})
\end{equation}

For the $\bar{B}^{0}_{s}\rightarrow f_{0}f_{0}(\sigma\sigma)$ decay, based on the mixing scheme the decay amplitude can be written as:
\begin{equation}
\begin{split}
\sqrt{2}{\cal{A}}(\bar{B}^{0}_{s}\rightarrow f_{0}f_{0})&=\sin^2{\theta}{\cal{A}}(\bar{B}^{0}_{s}\rightarrow f_{n}f_{n})
+\sin{2\theta}{\cal{A}}(\bar{B}^{0}_{s}\rightarrow f_{n}f_{s})+\cos^2{\theta}{\cal{A}}(\bar{B}^{0}_{s}\rightarrow f_{s}f_{s}),\\
\sqrt{2}{\cal{A}}(\bar{B}^{0}_{s}\rightarrow \sigma\sigma)&=\cos^2{\theta}{\cal{A}}(\bar{B}^{0}_{s}\rightarrow f_{n}f_{n})
-\sin{2\theta}{\cal{A}}(\bar{B}^{0}_{s}\rightarrow f_{n}f_{s})+\sin^2{\theta}{\cal{A}}(\bar{B}^{0}_{s}\rightarrow f_{s}f_{s}).
 \label{mixng}
\end{split}
\end{equation}
with
\begin{equation}
\begin{split}
{\cal{A}}(\bar{B}^{0}_{s}\rightarrow f_{s}f_{s})&=-2V_{tb}V^{*}_{ts}[(a_{3}+a_{4}+a_{5}-\frac{1}{2}a_{7}-\frac{1}{2}a_{9}-\frac{1}{2}a_{10})f_{B}{\emph M}^{LL}_{e}+(a_{6}-\frac{1}{2}a_{8})(\emph{F}^{SP}_{a}\bar{f}_{S}+\emph{F}^{SP}_{e}f_{B})\\
&+(C_{3}+C_{4}-\frac{1}{2}C_{9}-\frac{1}{2}C_{10})({\emph M}^{LL}_{c}+
{\emph M}^{LL}_{g})+(C_{5}-\frac{1}{2}C_{7})({\emph M}^{LR}_{c}+{\emph M}^{LR}_{g})+(C_{6}-\frac{1}{2}C_{8})({\emph M}^{SP}_{c}+{\emph M}^{SP}_{g})]
 \label{ampf}
\end{split}
\end{equation}
\begin{equation}
\begin{split}
\sqrt{2}{\cal{A}}(\bar{B}^{0}_{s}\rightarrow f_{n}f_{s})&=-V_{tb}V^{*}_{ts}[(C_{4}-\frac{1}{2}C_{10}){\emph M}^{LL}_{c}+(C_{6}-\frac{1}{2}C_{8}){\emph
M}^{SP}_{c}]
 \end{split}
 \end{equation}
and the decay
amplitude of the $\bar{B}^{0}_{s}\rightarrow f_{n}f_{n}$ is same to the $\bar{B}^{0}_{s}\rightarrow a_{0}a_{0}$ decays.
For the considered decay modes, the corresponding decay width is
\begin{equation}
{\Gamma}(\bar{B}^{0}_{s}\rightarrow S S)=\frac{G^2_{F}  m^3_{B_{s}}}{128 \pi}(1-2r_{S}^2)|{\cal{A}}(\bar{B}^{0}_{s}\rightarrow S S)|^2.
\end{equation}

Here, it is noticeable that the contribution from the factorizable annihilation diagrams in the $B^{0}_{s}\rightarrow a_{0} a_{0}$ decay is very small and can be safely neglected due to the isospin symmetry. And owing to the decay constant of the scalar meson $f_S=0$, we negelect all the responding
contribution in our calculation.

%%%--=================================================================
%%%=====           Numerical evaluation and discussions   ============
%%5===================================================================

\section{Numerical Results And Discussions}\label{sec:numer}
In this section, we will calculate the CP-averaged branching ratios and CP-violation asymmetries for the $\bar{B}^{0}_{s}\rightarrow SS$ decays and make some analyses about the results. First, we list the input parameters that are used in the calculations below. The masses and decay constant of the mesons, the lifetime of the $B_{s}$ are~\cite{Tanabashi:2018oca, Cheng:2005ye, Liu:2019ymi}
\begin{equation}
\begin{split}
       m_{B_{s}}&=5.367 {\rm GeV},   \quad  \bar m_{b}(\bar{m}_{b})=4.2 {\rm GeV},   \quad   m_{a_{0}}=0.98\pm0.02 {\rm GeV},\\ m_{f_{0}}&=0.99\pm0.02 {\rm GeV}, \quad f_{B_{s}}=227.2 \pm 3.4 {\rm MeV},   \quad  \tau_{B_{s}}=1.509 {\rm ps},\\
       m_{f_{n}}&=0.99 {\rm GeV},   \quad    m_{f_{s}}=1.02   {\rm  GeV}, \quad    m_{\sigma}=0.5   {\rm  GeV}.
\end{split}
\end{equation}
and in the CKM matrix elements, the involved Wolfenstein parameters are
\begin{equation}
\begin{split}
\lambda&=0.22453\pm0.00044,  \quad          \emph{A}=0.836\pm0.015, \\
\bar{\rho}&=0.122^{+0.018}_{-0.017}, \quad  \bar{\eta} =0.355^{+0.012}_{-0.011}.
\end{split}
\end{equation}
with the relations $\bar{\rho}={\rho}(1-\frac{\lambda^2}{2})$ and $\bar{\eta}={\eta}(1-\frac{\lambda^2}{2})$.

\subsection{Branching Ratios}

In this section, we separately give the results of the three considered decays $B^{0}_{s}\rightarrow a_{0}a_{0}$, $B^{0}_{s}\rightarrow f_{0}f_{0}$ and
$B^{0}_{s}\rightarrow \sigma\sigma$. For the $B^{0}_{s}\rightarrow a_{0}a_{0}$, this decay mode have both tree operators and penguin operators in the quark level. In SM, the $\gamma$ angle is associated with the CKM matrix element $V_{ub}$, which have the relationship $V_{ub}\simeq|V_{ub}|e^{-i\gamma}$. So we can leave the the CKM phase angle $\gamma$ as an unknown parameter, and write the decay amplitude of the $\bar{B}^{0}_{s}\rightarrow a_{0}a_{0}$ decay as
\begin{equation}
\begin{split}
\bar{\cal{A}}=V_{ub}V^*_{us}T-V_{tb}V^{*}_{ts}P=V_{ub}V^*_{us}T(1+z e^{i(\delta+\gamma)}),
\end{split}
\label{acharge}
\end{equation}
where the ratio $z=|V_{tb}V^{*}_{ts}/V_{ub}V^*_{us}|\cdot|P/T|$, and $\delta$ is the relative strong phase between the tree amplitudes$(T)$ and penguin amplitudes$(P)$. The value of $z$ and $\delta$ can be calculated from the pQCD.

Meanwhile, the decay amplitude of the conjugated decay mode ${B^{0}_{s}}\rightarrow a_{0}a_{0}$ can be written by replacing $V_{ub}V^*_{us}$ with $V^*_{ub}V_{us}$ and $V_{tb}V^{*}_{ts}$ with $V^{*}_{tb}V_{ts}$ as
\begin{equation}
\begin{split}
{\cal{A}}=V^*_{ub}V_{us}T-V^{*}_{tb}V_{ts}P=V^*_{ub}V_{us}T(1+z e^{i(\delta-\gamma)}).
\end{split}
\label{acon}
\end{equation}
Then from Eq.~(\ref{acharge}) and (\ref{acon}), the CP-averaged decay width of $\bar{B}^{0}_{s}({B^{0}_{s}})\rightarrow a^{+}_{0}a^{-}_{0}$ is
\begin{equation}
\begin{split}
\Gamma(\bar{B}^{0}_{s}({B^{0}_{s}})\rightarrow a^{+}_{0}a^{-}_{0}) &=\frac{G^2_{F}  m^3_{B_{s}}}{256 \pi}(1-2r^2_{a_{0}})(|{\cal A}|^2+|\overline{{\cal A}}|^2)\\
&=\frac{G^2_{F}  m^3_{B_{s}}}{128 \pi}(1-2r^2_{a_{0}})|V_{ub}^*V_{us}T|^2(1+2z\cos(\gamma)\cos(\delta)+z^2).
\label{CP-average}
\end{split}
\end{equation}

\begin{figure}[htbp]
 \centering
 \begin{tabular}{l}
 \includegraphics[width=0.5\textwidth]{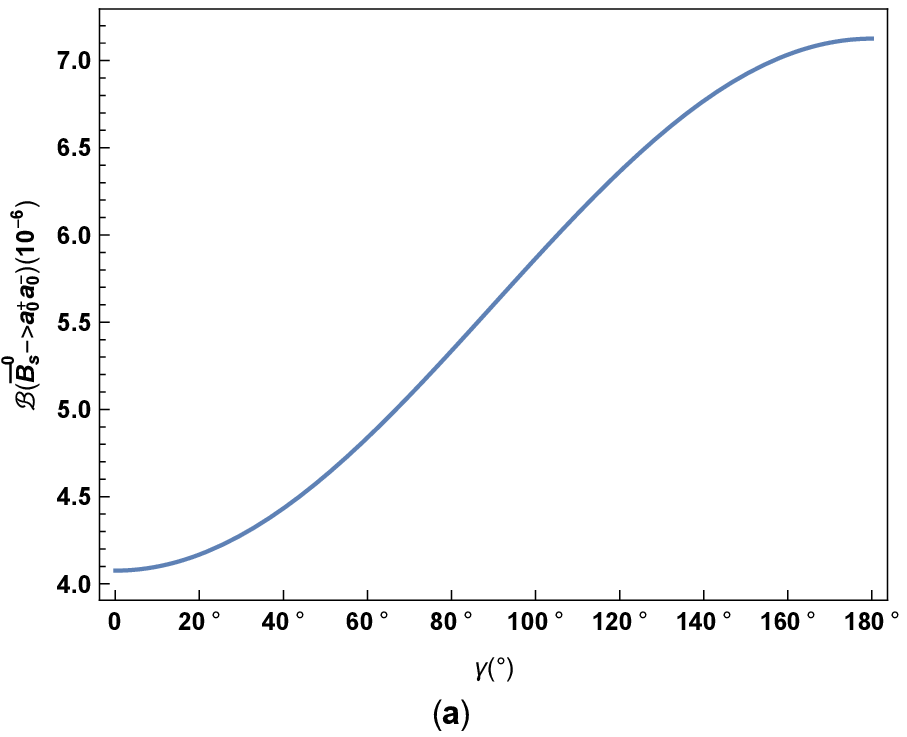}
 \includegraphics[width=0.5\textwidth]{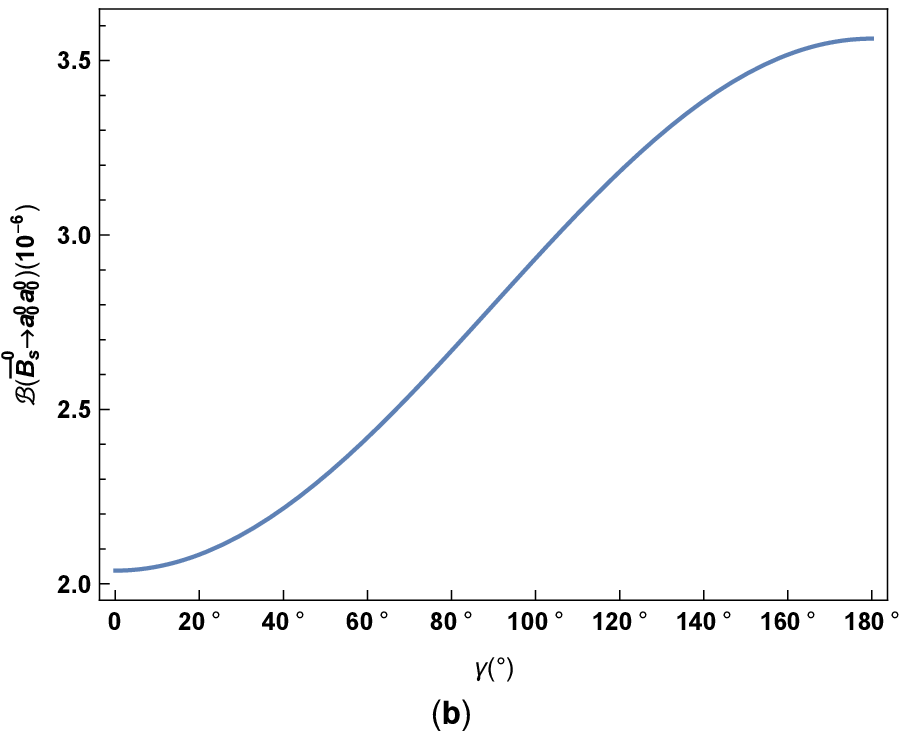}
 \end{tabular}
 \caption {(a)The branching ratio of the $\bar{B}^{0}_{s}\rightarrow a^{+}_{0}a^{-}_{0}$ decay as a function of $\gamma$;(b)The branching ratio of the $\bar{B}^{0}_{s}\rightarrow a^{0}_{0}a^{0}_{0}$ decay as a function of $\gamma$.}
   \label{S}
 \end{figure}

In Fig.~\ref{S}, we plot the average branching ratio of the decay $\bar{B}^{0}_{s}\rightarrow a^{+}_{0}a^{-}_{0}$ and $\bar{B}^{0}_{s}\rightarrow a^{0}_{0}a^{0}_{0}$ about the parameter $\gamma$ respectively. Since the CKM angle $\gamma$ is constrained as $\gamma$ around $73.5^{\circ}$ in Review of Particle Physics~\cite{Tanabashi:2018oca},
\begin{equation}
\gamma=(73.5^{+4.2}_{-5.1})^{\circ}
\end{equation}
we get from Fig.~\ref{S} when we take $\gamma$ as $70^{\circ}\sim 80^{\circ}$,
\begin{equation}
5.08\times 10^{-6} < {\cal B}(\bar{B}^{0}_{s}\rightarrow a^{+}_{0}a^{-}_{0}) < 5.34\times 10^{-6};
\end{equation}
\begin{equation}
2.54\times 10^{-6} < {\cal B}(\bar{B}^{0}_{s}\rightarrow a^{0}_{0}a^{0}_{0}) < 2.67\times 10^{-6}.
\end{equation}
The value of $z=6.67$ indicate that the amplitude of the penguin diagrams is almost $6.67$ times of that of tree diagrams. Therefore the main contribution come from the penguin diagrams in this decays, which enhance the results of the branching ratios.

When we utilize the input parameters and decay amplitudes,  furthermore leave the phase angle $\gamma$ aside, it is easy to get the CP-average branching ratios for both containing the charged and neutral scalar mesons decay modes, which are
\begin{equation}
{\cal B}(\bar{B}^{0}_{s}\rightarrow a^{+}_{0}a^{-}_{0})=5.17^{+1.62}_{-1.39}(B_{1})^{+0.24}_{-0.09}(B_{3})^{+1.23}_{-1.03}(\bar{f}_{a_{0}})^{+0.63}_{-0.55}(\omega_{b})^{+0.99}_{-0.67}(t_{i})\times10^{-6},
\end{equation}
\begin{equation}
{\cal B}(\bar{B}^{0}_{s}\rightarrow a^{0}_{0}a^{0}_{0})=2.58^{+0.81}_{-0.63}(B_{1})^{+0.12}_{-0.04}(B_{3})^{+0.62}_{-0.52}(\bar{f}_{a_{0}})^{+0.31}_{-0.27}(\omega_{b})^{+0.50}_{-0.33}(t_{i})\times10^{-6}.
\end{equation}

In pQCD approach, the wave functions of the initial and final mesons, whose are universal and channel independent, are the dominant inputs and have an important influence on the numerical results.
As it has been shown above, the primary errors come from the uncertainties of Gegenbauer moments $B_{1}=-0.93\pm 0.10$ and $B_{3}=0.14\pm 0.08$, the scalar decay constant $\bar{f}_{a_{0}}=0.365 \pm 0.020 {\rm GeV}$, the shape parameter $\omega_{b}=0.50\pm 0.05$ and the hard scale $t_{i}$, respectively.  The hard scale $t_{i}$ varies from $0.8t\sim1.2t$ (not changing $1/\emph{b}_{i}$, $i=1,2,3$), which characterizes the size of the next-leading-order contribution. The errors from the other uncertainties, such as the mass of the $m_{a_{0}}$ and CKM matrix elements, turn out to be small and can be neglected. It is apparent that the main errors are caused by the non-perturbative input parameters, which we need more precise experimental data to determine. By adding all of these vital uncertainties in quadrature, we get ${\cal B}(\bar{B}^{0}_{s}\rightarrow a^{+}_{0}a^{-}_{0})=(5.17^{+2.36}_{-1.94}) \times10^{-6}$ and ${\cal B}(\bar{B}^{0}_{s}\rightarrow a^{0}_{0}a^{0}_{0})=(2.58 ^{+1.18}_{-0.92})\times10^{-6}$.

In our previous work of $B^{0}_{s}\rightarrow \pi^{+} \pi^{-} $~\cite{Li:2004ep}(one of
the author have recalculated the $B^{0}_{s}\rightarrow \pi^{+} \pi^{-} $ and $B^{0}\rightarrow K^{+} K^{-} $ in $2012$~\cite{Xiao:2011tx}),
the theoretical results of these two decay modes are ${\cal B}(B^{0}_{s}\rightarrow \pi^{+} \pi^{-})=5.10\times 10^{-7}$
and ${\cal B}(B^{0}\rightarrow K^{+} K^{-})=1.56\times 10^{-7}$, where the corresponding experimental results about the branching ratios~\cite{Aaltonen:2011jv, Aaij:2016elb} of
these two decay modes approximately at the order of the $10^{-7} \sim 10^{-8}$. The predicted results of $\bar{B}^{0}_{s}\rightarrow a_{0}a_{0}$ for
both charged and neutral $a_{0}$ mesons, however, are at the order of $10^{-6}$ although these decay modes have the same quark components for both
initial and final state mesons and the only pure annihilation contributions.
So this results push us to make some comments about why the branching ratio of the $\bar{B}^{0}_{s}\rightarrow a^{+}_{0}a^{-}_{0}$ is more
large than the results of the $B^{0}_{s}\rightarrow \pi^{+} \pi^{-} $ decay and $B^{0}\rightarrow K^{+} K^{-} $ decay. By comparison, we can first
find that the main underlying reason is that the QCD dynamics of the scalar meson $a_{0}$ is different from that of the pseudoscalar meson $\pi$ and $K$, where at the leading twist the scalar meson $a_{0}$ is dominated by the odd Gegenbauer polynomials but the pseudoscalar mesons both $\pi$ and $K$
 are governed by the even Gegenbauer polynomials. Second the decay constant $\bar{f}_{a_{0}}$ is about two times than the decay constants
 of the $f_{\pi}$ and $f_{K}$ ~\cite{Xiao:2011tx, Nakamura:2010zzi}. These two reasons lead to the non-factorizable annihilation contribution
 is more large in the $\bar{B}^{0}_{s}\rightarrow a_{0}a_{0}$ mode. In Tab.~\ref{tab}, we list the decay amplitudes of the $\bar{B}^{0}_{s}\rightarrow a_{0}a_{0}$ for different distribution amplitudes of twist-2 or twist-3, and also
 we list the results of Ref.~\cite{Xiao:2011tx} about the decay mode $B^{0}\rightarrow K^{+} K^{-} $ for contrast.
 From Tab.~\ref{tab}, it is obvious that the twist-2 DA make dominant contribution, and the decay amplitudes of the $\bar{B}^{0}_{s}\rightarrow a_{0}a_{0}$ decay is approximately one order of the magnitude larger than that of the $B^{0}\rightarrow K^{+} K^{-} $.

\begin{table}[htbp]
\centering
\caption{The different source of twist-2 and twist-3 contribution.}
\label{tab}
\begin{tabular*}{\columnwidth}{@{\extracolsep{\fill}}lllll@{}}
\hline
\\
decay mode                                          &twist-2 $\phi_{a_{0}} (\phi^{A}_K)$        &twist-3   $\phi^S_{a_{0}} (\phi^{P}_K)$       &twist-3 $\phi^T_{a_{0}} (\phi^{T}_K)$\\
\hline
\\
${\cal A}(\bar{B}^{0}_{s}\rightarrow a^{+}_{0}a^{-}_{0})$     &$(-2.0-2.1 $i$)\times 10^{-4}$      & $(+4.2+4.1$i$)\times 10^{-5}$   & $(-2.27-0.79$i$)\times 10^{-6}$          \\

                                                   \\
${\cal A}(B^{0}\rightarrow K^{+} K^{-})$~\cite{Xiao:2011tx}   &$(-0.31-2.2$i$)\times 10^{-5}$     &$(-0.61-0.55$i$)\times 10^{-5}$     &$(-0.06-0.27$i$)\times 10^{-5}$\\
 \hline
\end{tabular*}
\end{table}

\begin{figure}[htbp]
 \centering
 \begin{tabular}{l}
 \includegraphics[width=0.5\textwidth]{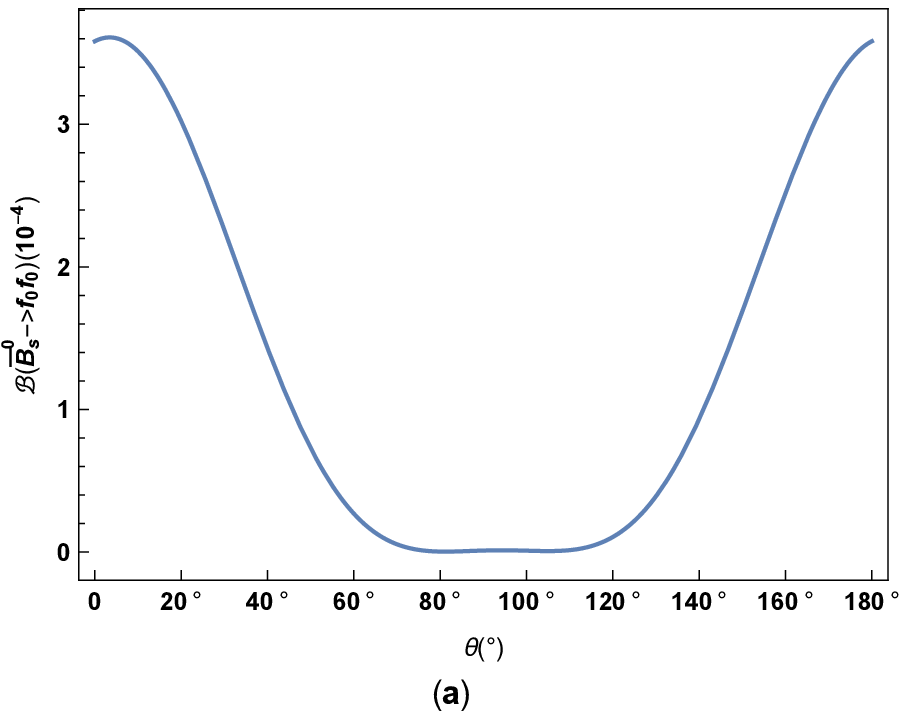}
 \includegraphics[width=0.5\textwidth]{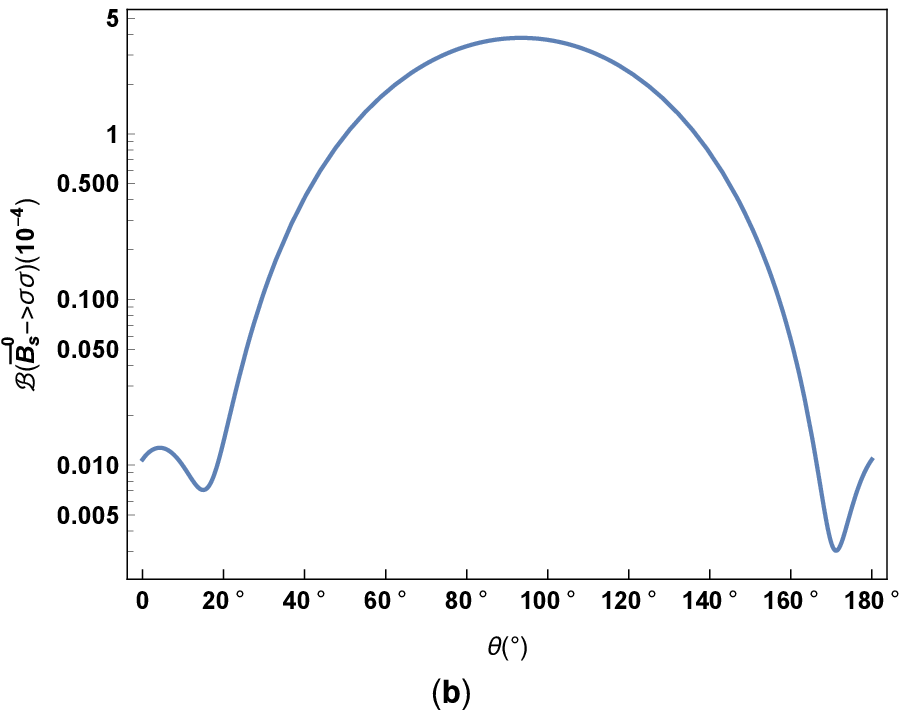}
 \end{tabular}
 \caption {(a)The branching ratio of the $\bar{B}^{0}_{s}\rightarrow f_{0}f_{0}$ decay as a function of mixing angle $\theta$;(b)The branching ratio of the
 $\bar{B}^{0}_{s}\rightarrow \sigma\sigma$ decay as a function of mixing angle $\theta$.}
   \label{980-500 mixing}
 \end{figure}

For the $\bar{B}^{0}_{s}\rightarrow f_{0}f_{0}$ decay, it is governed by the $b\rightarrow ss\bar{s}$ when we regard $f_0$ as the $s\bar{s}$, and this type decay only have the penguin operators due to the fact that the tree operators are forbidden. When introducing the mixing effect from the component of
the$(u\bar{u}+d\bar{d})/\sqrt{2}$, we take the mixing angle $\theta$ as a free parameter, and then plot the branching ratio's dependence on the mixing angle in
Fig.~\ref{980-500 mixing}. If the $f_{0}$ is the pure $s\bar{s}$ component, namely the mixing angle $\theta=0^\circ$, the branching ratio of the $\bar{B}^{0}_{s}\rightarrow
f_{0}f_{0}$ is approximately $3.6\times 10^{-4}$, and when including the mixing effect of the $(u\bar{u}+d\bar{d})/\sqrt{2}$, the result change clearly
which we can read from Fig.~\ref{980-500 mixing}(a).
For the $\bar{B}^{0}_{s}\rightarrow \sigma\sigma$ decay, there are still a lot of uncertainties about the wave function of $\sigma$ meson, we choose the same decay
constant for $f_{n}$ and $f_{s}$ in our calculations, just as it has been done in Ref.~\cite{Cheng:2005nb}. The results of this decay is contrary to
the $\bar{B}^{0}_{s}\rightarrow f_{0}f_{0}$, which is dominated by the $\sin$ law that we just see from the Eq.~(\ref{mixng}), when taking the mixing
angle $\theta=0^\circ$, the branching ratio of this decay is very small, and it will increase about one or two
magnitude
in consideration of the mixing effect of the $s\bar{s}$. The decay amplitude of the $\bar{B}^{0}_{s}\rightarrow f_{0}f_{0}(\sigma\sigma)$ contain three parts,
$f_{n}f_{n}$, $f_{n}f_{s}$ and $f_{s}f_{s}$ , and the main contribution comes from $f_{s}f_{s}$. The oscillation
near the two ends of the $\theta$-coordinate in Fig.~\ref{980-500 mixing}(b) mainly due to the interference from $\bar{B}^{0}_{s}\rightarrow f_{n}f_{n}$
and its contribution obey the $\cos$ law for $\bar{B}^{0}_{s}\rightarrow \sigma\sigma$ decay that
will obviously enhance the two ends of theta axis in Fig.~\ref{980-500 mixing}(b). Taking both the two decays into account, we can find that the mixing angle can be constrained in the range $[19^\circ, 66^\circ]$ and $[119^\circ, 166^\circ]$ because it will be nearly zero when taking other values, and if combining the
known results that obtained from the experiment, the range will be smaller. The mixing angle range that we get are also consistent with the data of the Ref.~\cite{ Cheng:2002ai, Anisovich:2001zp, Gokalp:2004ny,Anisovich:2002wy}.

The mixing angle is not clear up to now, and there are a lot of works to constrain the angle range. The LHCb Collaboration firstly announced the upper limit $|\theta|<31^\circ$ for the mixing angle of the $\sigma-f_{0}$ in Ref.~\cite{Aaij:2013zpt}. So we set the two value $\theta=25^\circ$ and $\theta=30^\circ$ to make some calculation respectively, the branching ratios are presented as

(1) $\theta=25^\circ$
 \begin{equation}
 \begin{split}
 {\cal B}(\bar{B}^{0}_{s}\rightarrow
 f_{0}f_{0})&=2.66^{+0.19}_{-0.18}(B_{1})^{+0.31}_{-0.29}(B_{3})^{+0.63}_{-0.53}(\bar{f}_{S})^{+0.32}_{-0.27}(\omega_{b})^{+0.73}_{-0.50}(t_{i})\times10^{-4},\\
 &=2.66^{+1.08}_{-0.85}\times10^{-4};\\
 {\cal B}(\bar{B}^{0}_{s}\rightarrow
 \sigma\sigma)&=4.35^{+0.22}_{-0.14}(B_{1})^{+0.41}_{-0.37}(B_{3})^{+0.52}_{-0.87}(\bar{f}_{S})^{+1.00}_{-0.83}(\omega_{b})^{+1.25}_{-0.80}(t_{i})\times10^{-6}\\
 &=4.35^{+1.75}_{-1.50}\times10^{-6}.
 \label{25}
 \end{split}
 \end{equation}
(2) $\theta=30^\circ$
\begin{equation}
\begin{split}
{\cal B}(\bar{B}^{0}_{s}\rightarrow f_{0}f_{0})&=2.26^{+0.16}_{-0.16}(B_{1})^{+0.26}_{-0.26}(B_{3})^{+0.53}_{-0.45}(\bar{f}_{S})^{+0.26}_{-0.23}(\omega_{b})^{+0.61}_{-0.42}(t_{i})\times10^{-4}\\
&=2.26^{+0.90}_{-0.72}\times10^{-4},\\
{\cal B}(\bar{B}^{0}_{s}\rightarrow \sigma\sigma)&=1.11^{+0.04}_{-0.04}(B_{1})^{+0.01}_{-0.02}(B_{3})^{+0.21}_{-0.22}(\bar{f}_{S})^{+0.21}_{-0.18}(\omega_{b})^{+0.32}_{-0.21}(t_{i})\times10^{-5}\\
&=1.11^{+0.44}_{-0.36}\times10^{-5}.
\label{30}
\end{split}
\end{equation}
We can get the same results when the value of $\theta$ are close to the $161^\circ$ and $157^\circ$, respectively. In every second line of the Eq.~(\ref{25}) and Eq.~(\ref{30}), the theoretical errors that we considered are added in quadrature. The main reason for the branching ratio of $\bar{B}^{0}_{s}\rightarrow f_{0}f_{0}$ is larger than that of $\bar{B}^{0}_{s}\rightarrow \sigma\sigma$ is that the mass of $f_{0}$ is almost one time heavier than that of $\sigma$.

For the mixing of $a^0_0-f_{0}$, we directly take the mixing intensity $\xi_{fa}$,
\begin{equation}
\begin{split}
&\xi_{fa}=(0.99\pm0.16\pm0.30\pm0.19)\times 10^{-2} \quad  \mathrm{(solution \quad \uppercase\expandafter{\romannumeral1})},\\
&\xi_{fa}=(0.41\pm0.13\pm0.17\pm0.13)\times 10^{-2} \quad  \mathrm{(solution \quad \uppercase\expandafter{\romannumeral2})}.
\end{split}
\end{equation}
which are first measured in the BES III collaboration~\cite{Ablikim:2018pik}, and the relation $|\xi_{fa}|\simeq \tan^2 \phi$ is applied to get the mixing angle $\phi$~\cite{Aliev:2018bln}.
\begin{equation}
\begin{split}
&\phi=(5.45\pm1.65)^{\circ} \quad  \mathrm{(sloution \quad \uppercase\expandafter{\romannumeral1})},\\
&\phi=(3.02\pm2.21)^{\circ} \quad  \mathrm{(sloution \quad \uppercase\expandafter{\romannumeral2})}.
\end{split}
\end{equation}
From the value, we can conclude that the mixing angle is so small that it will not change our results largely.

Here we also make some comments when the final state of the decay mode treated as the four-quark structure. As we mentioned in the introduction, there is an open problem that the inner structure of the scalar meson are not well identified. In this work, we regard $a_0$, $f_0$ and $\sigma$ as the $q\bar{q}$ in the traditional quark model and make some calculations within the perturbative QCD approach. But when we want to make some predictions of the tetraquark picture in the perturbative QCD approach, we can not make directly computations because we do not known the necessary physical quantities, such as the wave function of the scalar mesons of four-quark picture. However, we can image a picture is that the other $q\bar{q}$ pairs must be extracted from the sea quarks when the scalar mesons are four-quark state, and it would be expected that the branching ratios of these decay modes in tetraquark picture are smaller than that in two-quark model.

\subsection{CP Violation Parameters}

Now, we will calculate the CP violation parameters of the $\bar{B}^{0}_{s}\rightarrow a_{0}a_{0}$ decays in this subsection. The CP violation parameters of the $\bar{B}^{0}_{s}\rightarrow a_{0}a_{0}$ for both charged and neutral $a_{0}$ mesons are same because the decay amplitude of these two decay modes are similar and the factor in the front of the decay width formula can be reduced. In SM, CP violation originated from the CKM weak angle. For the neutral $B_{s}^{0}$ meson decays, we should take the effect of $\bar{B}^{0}_{s}-{B^{0}_{s}}$ mixing into account, and the time dependent CP violation parameters of the two $\bar{B}^{0}_{s}\rightarrow a_{0}a_{0}$ decays with charged and neutral scalar mesons can be defined as
\begin{equation}
\begin{split}
A_{\rm CP}& = \frac{
\Gamma\left(B^0_{s}(\Delta t) \to a_{0}a_{0}\right )-\Gamma\left
(\bar{B}^0_{s}(\Delta t) \to a_{0}a_{0}\right)}{ \Gamma\left
(B^0_{s}(\Delta t) \to a_{0}a_{0}\right ) + \Gamma\left
(\bar{B}^0_{s}(\Delta t) \to a_{0}a_{0}\right ) }\non
&= A_{\rm CP}^{\rm dir} \cos(\Delta m  \Delta t) + A_{\rm CP}^{\rm mix} \sin (\Delta m \Delta t),
\label{acp-def}
\end{split}
\end{equation}
where $\Delta m$ is the mass difference between the two neutral $B^{0}_{s}$($\bar{B}^{0}_{s}$) mass eigenstates, and $\Delta t= t_{CP}-t_{tag}$ is the time difference between the tagged $B^{0}_{s}$($\bar{B}^{0}_{s}$) and the accompanying $\bar{B}^{0}_{s}$($B^{0}_{s}$) with opposite $b$ flavor decaying to the final $CP$ eigenstate $a_{0}a_{0}$ at the time $t_{CP}$.

From Eqs.~(\ref{acharge}) and (\ref{acon}), the direct CP violation parameter $A_{\rm CP}^{\rm dir}$ can be parameterized as
\begin{equation}
\begin{split}
A_{\rm CP}^{\rm dir}&=\frac{|A|^2-|\bar{A}|^2}{|A|^2+|\bar{A}|^2}=\frac{2z\sin(\delta) \sin(\gamma)}{1+2z\cos(\delta) \cos(\gamma)+z^2}.
\label{CP-dir}
\end{split}
\end{equation}

\begin{figure}[htbp]
 \centering
 \begin{tabular}{l}
 \includegraphics[width=0.5\textwidth]{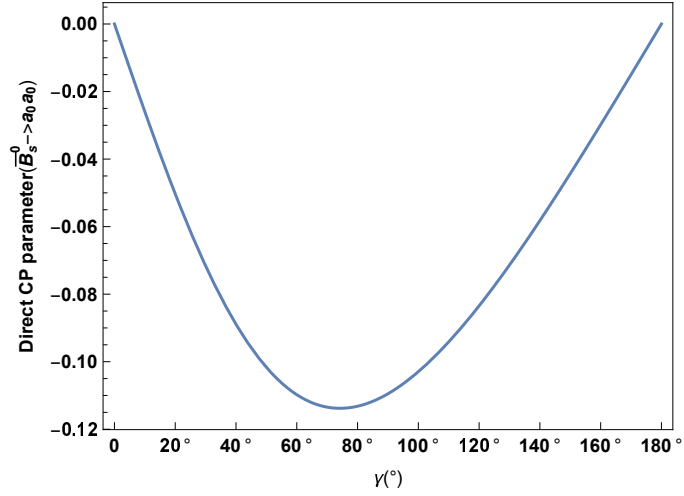}
 \end{tabular}
 \caption {The direct CP violation parameter of the $\bar{B}^{0}_{s}(B^{0}_{s})\rightarrow a_{0}a_{0}$ decay as a function of $\gamma$.}
   \label{CPd}
 \end{figure}

It is obvious that the $A_{\rm CP}^{\rm dir}$ is approximately proportional to CKM angle $\sin(\gamma)$,
strong phase $\sin(\delta)$, and the relative size $z$ between the
penguin contribution and tree contribution. We plot the direct CP violation parameter $A_{\rm CP}^{\rm dir}$ as the function of the weak angle $\gamma$ in Fig.~\ref{CPd}, and one can see that the $A_{\rm CP}^{\rm dir}$ is approximately $-11.4\%$ at the peak when the $\gamma$ is $70^{\circ}< \gamma < 80^{\circ}$. The relative small direct CP asymmetry is also a result of the main contributions coming from penguin diagrams in this decays.

The involved mixing-induced CP violation parameter $A_{\rm CP}^{\rm mix}$ can be written as
\begin{equation}
A_{\rm CP}^{\rm mix}=\frac{-2 {\rm Im}(\lambda_{CP})}{1+|\lambda_{CP}|^2},
\label{CP-mix}
\end{equation}
with the CP violation parameters $\lambda_{CP}$
\begin{equation}
\begin{split}
\lambda_{CP}=\eta_{CP} \frac{V^{*}_{tb}V_{ts}}{V_{tb}V^{*}_{ts}}\frac{\langle a_{0}a_{0}|H_{eff}|\bar{B}^0_{s}\rangle}{\langle a_{0}a_{0}|H_{eff}|{B}^0_{s}\rangle}
=e^{-2i\gamma}\frac{1+ze^{i(\delta+\gamma)}}{1+ze^{i(\delta-\gamma)}},
\end{split}
\end{equation}
in which $\eta_{CP}$ is the CP-eigenvalue of the final state.

\begin{figure}[htbp]
 \centering
 \begin{tabular}{l}
 \includegraphics[width=0.5\textwidth]{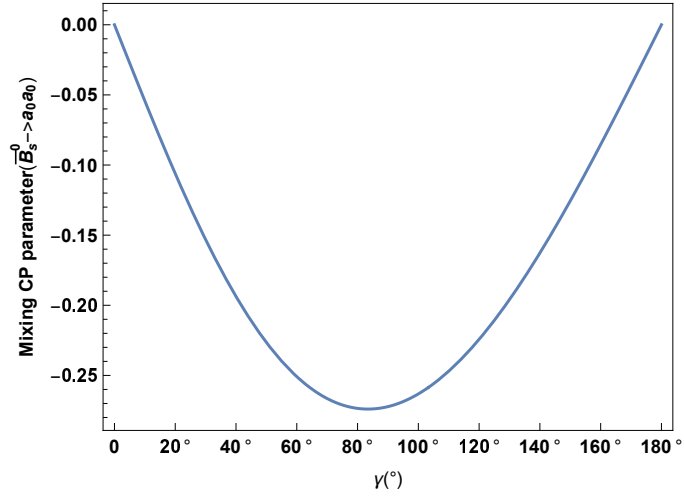}
 \end{tabular}
 \caption {The mixing CP violation parameter of the $\bar{B}^{0}_{s}(B^{0}_{s})\rightarrow a_{0}a_{0}$ decay as a function of $\gamma$.}
   \label{CPm}
 \end{figure}
If $z$ is a very small number, i. e., the penguin diagram contribution is suppressed comparing with the tree diagram contribution, the
mixing induced CP asymmetry parameter $A_{\rm CP}^{\rm mix}$ is proportional to $\sin2\gamma$, which will be a good place for the CKM
angle $\gamma$ measurement. However as we have already mentioned, $z$(=6.67) is large. We give the mixing CP asymmetry in Fig.~\ref{CPm}, one
can see that $A_{\rm CP}^{\rm mix}$ just
like the case of direct CP violation, it is almost symmetric
and the symmetry axis is near $\gamma=\pi/2$. It is close to $-27.0\%$ when the angle $\gamma$ is constrained as $\gamma$ around $73.5^{\circ}$. At present, there are
no CP asymmetry measurements in experiment but the possible large CP
violation we predict for $\bar{B}^{0}_{s}\rightarrow a_{0}a_{0}$ decays might be observed in the coming LHC-b experiments.

For the $\bar{B}^0_s\rightarrow f_{0}f_{0}$ decay, it is a pure penguin process when we regard $f_0$ as $s\bar{s}$ state and in this case, there is no weak phase that leads the direct CP violation parameter equal to zero. Furthermore, it is very small when take the mixing of the $(u\bar{u}+d\bar{d})/\sqrt{2}$
into account. For the $\bar{B}^0_s\rightarrow \sigma\sigma$ decay, it is a rare mode, the CKM matrix elements $|V_{us}V_{ub}|\ll |V_{ts}V_{tb}|$, which make the tree amplitudes are suppressed. From Eq.~(\ref{acp-def}), the direct and mixing CP asymmetries can be defined as follows:
\begin{equation}
\begin{split}
A_{\rm CP}^{\rm dir}=\frac{1-|\lambda_{CP}|^2}{1+|\lambda_{CP}|^2},
A_{\rm CP}^{\rm mix}=\frac{-2 {\rm Im}(\lambda_{CP})}{1+|\lambda_{CP}|^2},
\end{split}
\end{equation}

\begin{figure}[htbp]
\centering
\begin{tabular}{l}
\includegraphics[width=0.5\textwidth]{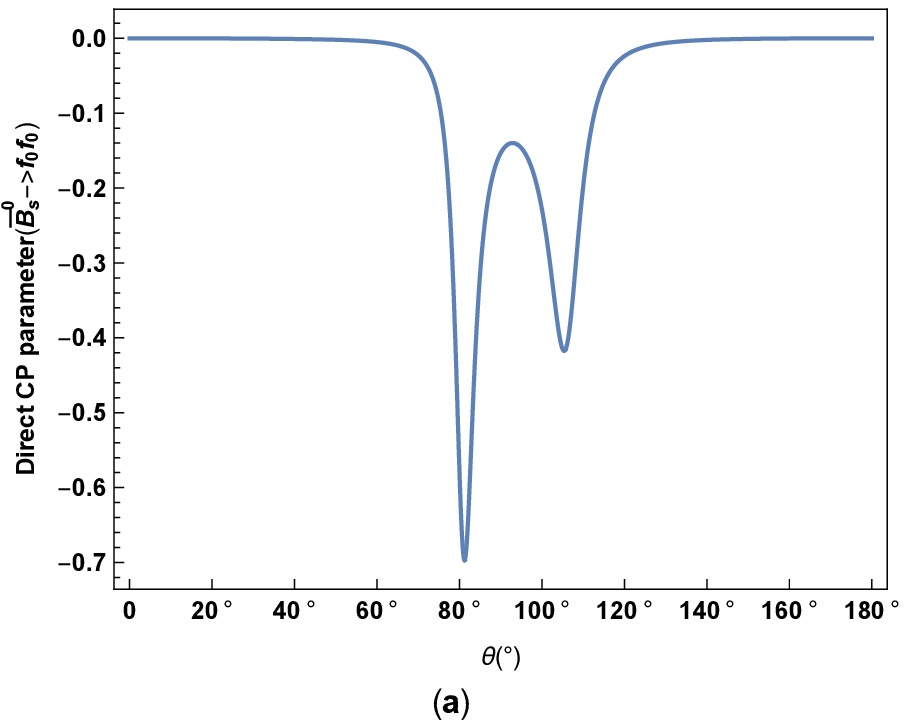}
\includegraphics[width=0.5\textwidth]{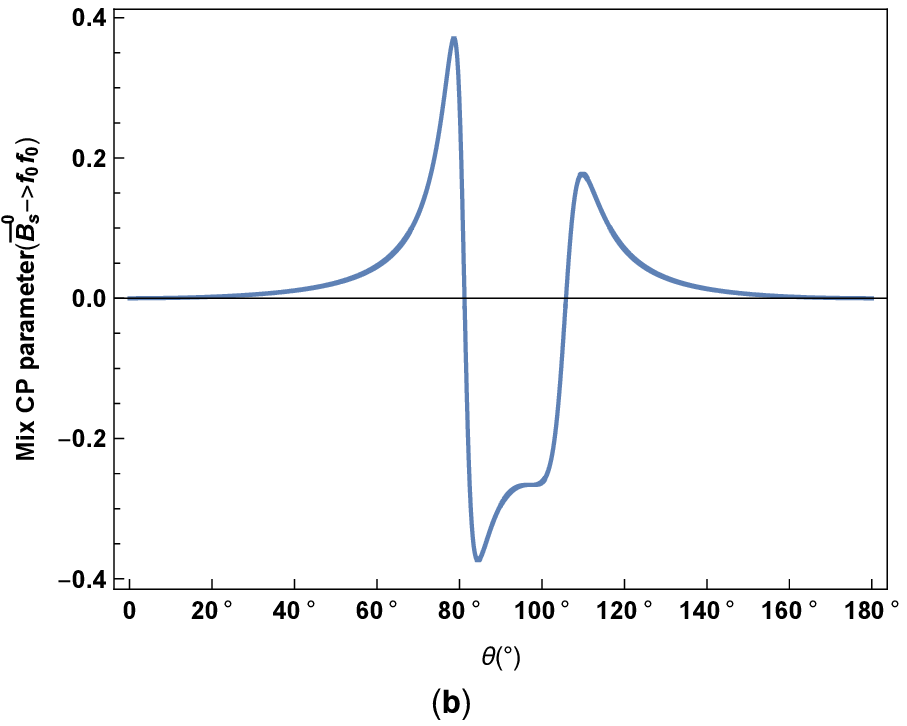}\\
\includegraphics[width=0.5\textwidth]{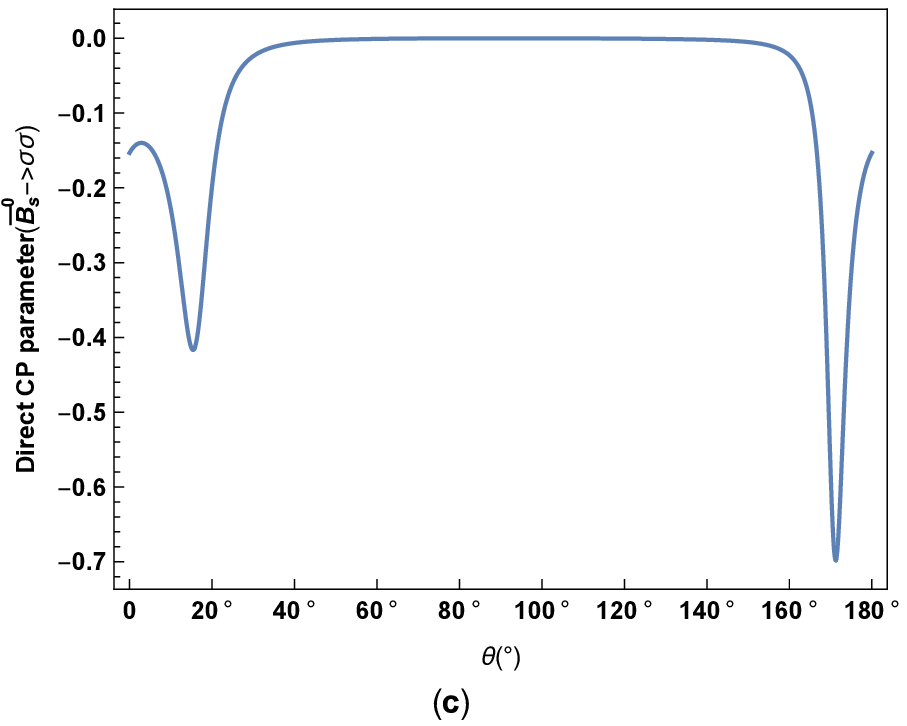}
\includegraphics[width=0.5\textwidth]{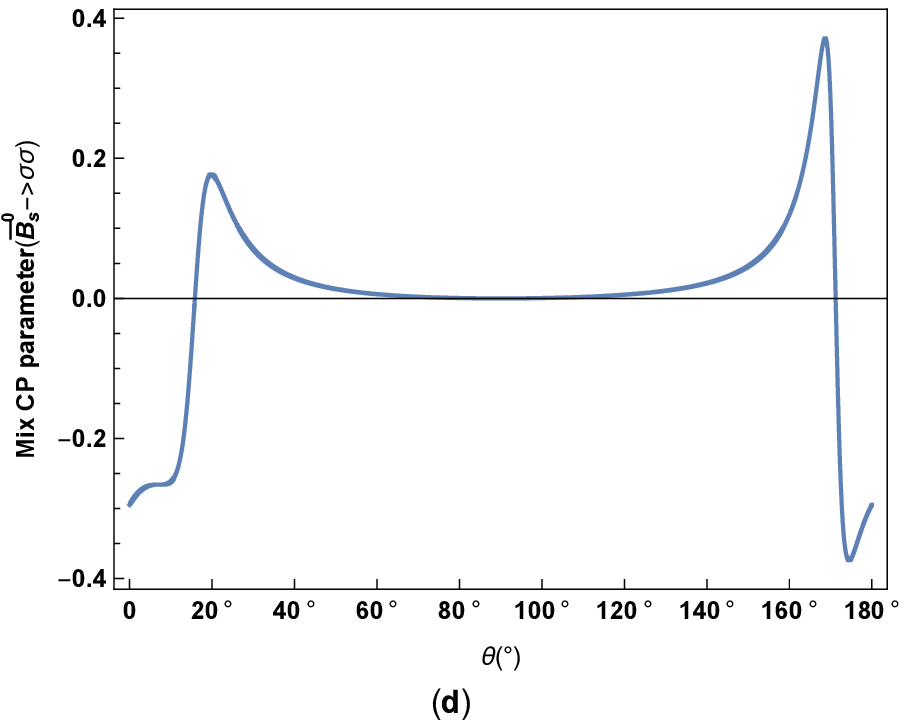}
\end{tabular}
\caption {The direct and mixing CP violation parameter of the $\bar{B}^{0}_{s}(B^{0}_{s})\rightarrow f_{0}f_{0}$ and $\bar{B}^{0}_{s}(B^{0}_{s})\rightarrow \sigma\sigma$ as a function of mixing angle $\theta$.}
\label{f980-500CP}
\end{figure}

Based on the mixing scheme, we give the CP asymmetries's dependence on the mixing angle $\theta$ in Fig.~\ref{f980-500CP}

Here, we use the same value of the $\theta=25^\circ$ to make some prediction,
\begin{equation}
\begin{split}
&A_{\rm CP}^{\rm dir}(\bar{B}^{0}_{s}\rightarrow f_{0}f_{0})=0, \\
&A_{\rm CP}^{\rm mix}(\bar{B}^{0}_{s}\rightarrow f_{0}f_{0})=0.3\%,   \\
&A_{\rm CP}^{\rm dir}(\bar{B}^{0}_{s}\rightarrow \sigma\sigma)=-6.0\%, \\
&A_{\rm CP}^{\rm mix}(\bar{B}^{0}_{s}\rightarrow \sigma\sigma)=11.7\%,   \\
\end{split}
\end{equation}
As for the $\bar{B}^{0}_{s}\rightarrow f_{0}f_{0}$, if we consider $f_0$ as a pure $s\bar{s}$ state, there is no CP violations; if we consider it as a mixing between $s\bar{s}$
and $q\bar{q}$, we find the interference has little influence on the CP violation parameters. Because the mixing angle can not be determined in a direct method, our results also can be used to constrain the range of the mixing angle $\theta$ if it were observed in the experiment.

\section{Summary} \label{sec:summary}

In this paper, we make predictions of the decay $\bar{B}^{0}_{s}\rightarrow SS(S=a_{0}(980), f_{0}(980,500))$ within the pQCD approach for the first time. Basing on the recently experimental results which provide a direct information about the constituent two-quark components in the corresponding $a_{0}$ wave function and the theoretical presentations of the scalar meson in Scenario~1, we calculate the branching ratios and CP violation parameters of the decay $\bar{B}^{0}_{s}\rightarrow a_{0}a_{0}$ for both charged and
neutral $a_{0}$ states and the decay $\bar{B}^{0}_{s}\rightarrow f_{0}(\sigma)f_{0}(\sigma)$. Our calculations show that:(1) the $\bar{B}^{0}_{s}\rightarrow a_{0}a_{0}$ decay modes have relative large branching ratios, which are ${\cal B}(\bar{B}^{0}_{s}\rightarrow a^{+}_{0}a^{-}_{0})=(5.17^{+2.36}_{-1.94}) \times10^{-6}$ and ${\cal B}(\bar{B}^{0}_{s}\rightarrow a^{0}_{0}a^{0}_{0})=(2.58 ^{+1.18}_{-0.92})\times10^{-6}$, and there is also large CP violation in the decay model;
(2) the branching fraction of $\bar{B}^{0}_{s}\rightarrow f_{0}(\sigma)f_{0}(\sigma)$ are at the order of the $10^{-4}$($10^{-6}$). Because the mixing angle can not be determined in a direct method, our results also can be used to constrain the range of the mixing angle $\theta$ if it were observed in the experiment.
In the end, we hope the results can be tested by the running LHC-b experiments in the near future, and, of course, it would help us to get a better understanding of the QCD behavior of the scalar mesons.

%%%--=================================================================
%%%=====            Acknowledgements        ==========================
%%5===================================================================

\section*{acknowledgments}

The authors would like to thank Dr. Ming-Zhen Zhou and Dr. Shan Cheng for some valuable discussions. This work is supported by the National Natural Science Foundation of China under Grant No.11047028 and No.11875226, and by the Fundamental Research Funds of the Central Universities, Grant Number XDJK2012C040.

\section*{Appendix}

In this part, we list some formulae that used in the above calculations.
The hard scattering kernels function $h_{i}(i=a, c, e, g)$ involved in the above expression are written as:
\begin{align}
&h^{1}_{a}(\emph{x}_{1}, \emph{x}_{2}, \emph{b}_{1}, \emph{b}_{2})=K_{0}(M_{B_{s}}\emph{b}_{1}\sqrt{\emph{x}_{1}(1-\emph{x}_{2})})
\times[\theta(b_{2}-b_{1})I_{0}(M_{B_{s}}\emph{b}_{1}\sqrt{1-\emph{x}_{2}})K_{0}(M_{B_{s}}\emph{b}_{2}\sqrt{1-\emph{x}_{2}})
+(\emph{b}_{2}\leftrightarrow \emph{b}_{1})],\\
&h^{2}_{a}(\emph{x}_{1}, \emph{x}_{2}, \emph{b}_{1}, \emph{b}_{2})=K_{0}(M_{B_{s}}\emph{b}_{2}\sqrt{\emph{x}_{1}(1-\emph{x}_{2})})
\times[\theta(b_{2}-b_{1})I_{0}(M_{B_{s}}\emph{b}_{1}\sqrt{\emph{x}_{1}})K_{0}(M_{B_{s}}\emph{b}_{2}\sqrt{\emph{x}_{1}})
+(\emph{b}_{2}\leftrightarrow \emph{b}_{1})],\\
&h^{1}_{c}(\emph{x}_{1}, \emph{x}_{2}, \emph{x}_{3}, \emph{b}_{2}, \emph{b}_{3})=[\theta(\emph{b}_{2}-\emph{b}_{3})I_{0}(M_{B_{s}}\emph{b}_{3}\sqrt{\emph{x}_{1}(1-\emph{x}_{2})})K_{0}(M_{B_{s}}\emph{b}_{2}\sqrt{\emph{x}_{1}(1-\emph{x}_{2})})+(\emph{b}_{2}\leftrightarrow \emph{b}_{3})]
\nonumber \nonumber\\ &\times
\begin{cases}
K_{0}(M_{B_{s}}\emph{b}_{3}\sqrt{\emph{x}_{1}+\emph{x}_{2}+\emph{x}_{3}-\emph{x}_{1}\emph{x}_{2}-\emph{x}_{2}\emph{x}_{3}-1}), & {\emph{x}_{1}+\emph{x}_{2}+\emph{x}_{3}-\emph{x}_{1}\emph{x}_{2}-\emph{x}_{2}\emph{x}_{3}-1 \geq 0}\\
\frac{\emph{i}{\pi}}{2} H^{(1)}_{0}(M_{B_{s}}\emph{b}_{3}\sqrt{|\emph{x}_{1}+\emph{x}_{2}+\emph{x}_{3}-\emph{x}_{1}\emph{x}_{2}-\emph{x}_{2}\emph{x}_{3}-1|}), & {\emph{x}_{1}+\emph{x}_{2}+\emph{x}_{3}-\emph{x}_{1}\emph{x}_{2}-\emph{x}_{2}\emph{x}_{3}-1 < 0}\\
\end{cases}\\
&h^{2}_{c}(\emph{x}_{1}, \emph{x}_{2}, \emph{x}_{3}, \emph{b}_{2}, \emph{b}_{3})=[\theta(\emph{b}_{2}-\emph{b}_{3})I_{0}(M_{B_{s}}\emph{b}_{3}\sqrt{\emph{x}_{1}(1-\emph{x}_{2})})K_{0}(M_{B_{s}}\emph{b}_{2}\sqrt{\emph{x}_{1}(1-\emph{x}_{2})})+(\emph{b}_{2}\leftrightarrow \emph{b}_{3})]
\nonumber \nonumber\\ &\times
\begin{cases}
K_{0}(M_{B_{s}}\emph{b}_{3}\sqrt{\emph{x}_{1}-\emph{x}_{3}-\emph{x}_{1}\emph{x}_{2}+\emph{x}_{2}\emph{x}_{3}}), & {\emph{x}_{1}-\emph{x}_{3}-\emph{x}_{1}\emph{x}_{2}+\emph{x}_{2}\emph{x}_{3} \geq 0}\\
\frac{\emph{i}{\pi}}{2} H^{(1)}_{0}(M_{B_{s}}\emph{b}_{3}\sqrt{|\emph{x}_{1}-\emph{x}_{3}-\emph{x}_{1}\emph{x}_{2}+\emph{x}_{2}\emph{x}_{3}|}), & {\emph{x}_{1}-\emph{x}_{3}-\emph{x}_{1}\emph{x}_{2}+\emph{x}_{2}\emph{x}_{3} < 0}\\
\end{cases}\\
&h^{1}_{e}(\emph{x}_{2}, \emph{x}_{3}, \emph{b}_{2}, \emph{b}_{3}) = \frac{\pi i}{2}H^{(1)}_{0}(M_{B_{s}}\emph{b}_{2}\sqrt{\emph{x}_{2}\emph{x}_{3}})
\times[\theta(\emph{b}_{2}-\emph{b}_{3})J_{0}(M_{B_{s}}\emph{b}_{3}\sqrt{\emph{x}_{3}})\frac{\pi i}{2}H^{(1)}_{0}(M_{B_{s}}\emph{b}_{2}\sqrt{\emph{x}_{3}})+(\emph{b}_{2}\leftrightarrow \emph{b}_{3})],\\
&h^{2}_{e}(\emph{x}_{2}, \emph{x}_{3}, \emph{b}_{2}, \emph{b}_{3})=\frac{\pi i}{2}H^{(1)}_{0}(M_{B_{s}}\emph{b}_{3}\sqrt{\emph{x}_{2}\emph{x}_{3}})
\times
[\theta(\emph{b}_{2}-\emph{b}_{3})J_{0}(M_{B_{s}}\emph{b}_{3}\sqrt{\emph{x}_{2}})\frac{\pi i}{2}H^{(1)}_{0}(M_{B_{s}}\emph{b}_{2}\sqrt{\emph{x}_{2}})+(\emph{b}_{2}\leftrightarrow \emph{b}_{3})],\\
&h^{1}_{g}(\emph{x}_{1}, \emph{x}_{2}, \emph{x}_{3}, \emph{b}_{1}, \emph{b}_{2})=[\theta(\emph{b}_{2}-\emph{b}_{1})J_{0}(M_{B_{s}}\emph{b}_{1}\sqrt{\emph{x}_{2}\emph{x}_{3}})\frac{\pi \emph{i}}{2}H^{(1)}_{0}(M_{B_{s}}\emph{b}_{2}\sqrt{\emph{x}_{2}\emph{x}_{3}})+(\emph{b}_{2}\leftrightarrow \emph{b}_{1})]
\nonumber \nonumber\\ &\times
\begin{cases}
K_{0}(M_{B_{s}}\emph{b}_{1}\sqrt{\emph{x}_{1}\emph{x}_{2}-\emph{x}_{2}\emph{x}_{3}}), & {\emph{x}_{1}\emph{x}_{2}-\emph{x}_{2}\emph{x}_{3} \geq 0}\\
\frac{\emph{i}{\pi}}{2} H^{(1)}_{0}(M_{B_{s}}\emph{b}_{1}\sqrt{|\emph{x}_{1}\emph{x}_{2}-\emph{x}_{2}\emph{x}_{3}|}), & {\emph{x}_{1}\emph{x}_{2}-\emph{x}_{2}\emph{x}_{3} < 0}\\
\end{cases}\\
&h^{2}_{g}(\emph{x}_{1}, \emph{x}_{2}, \emph{x}_{3}, \emph{b}_{1}, \emph{b}_{2})=[\theta(\emph{b}_{2}-\emph{b}_{1})J_{0}(M_{B_{s}}\emph{b}_{1}\sqrt{\emph{x}_{2}\emph{x}_{3}}) \frac{\pi \emph{i}}{2}H^{(1)}_{0}(M_{B_{s}}\emph{b}_{2}\sqrt{\emph{x}_{2}\emph{x}_{3}})+(\emph{b}_{2}\leftrightarrow \emph{b}_{1})]
\nonumber \nonumber \nonumber  \quad \quad \quad \quad \quad \quad \quad \quad \quad\\ &\times
\begin{cases}
K_{0}(M_{B_{s}}\emph{b}_{1}\sqrt{\emph{x}_{1}+\emph{x}_{2}+\emph{x}_{3}-\emph{x}_{1}\emph{x}_{2}-\emph{x}_{2}\emph{x}_{3}}), & {\emph{x}_{1}+\emph{x}_{2}+\emph{x}_{3}-\emph{x}_{1}\emph{x}_{2}-\emph{x}_{2}\emph{x}_{3} \geq 0}\\
\frac{\emph{i}{\pi}}{2} H^{(1)}_{0}(M_{B_{s}}\emph{b}_{1}\sqrt{|\emph{x}_{1}+\emph{x}_{2}+\emph{x}_{3}-\emph{x}_{1}\emph{x}_{2}-\emph{x}_{2}\emph{x}_{3}|}), & {\emph{x}_{1}+\emph{x}_{2}+\emph{x}_{3}-\emph{x}_{1}\emph{x}_{2}-\emph{x}_{2}\emph{x}_{3} < 0}
\end{cases}
\end{align}
where $J_{0}$ is the Bessel function and $K_{0}$, $I_{0}$ are modified Bessel function with $H^{(1)}_{0}(\emph{x})=J_{0}(\emph{x})+iY_{0}(\emph{x})$.

The evolution function $E(t_{i})$ is defined by
\begin{equation}
\begin{split}
&E_{ef}(t_{i})=\alpha_{s}(t_{i})\exp[-S_{B^{0}_{s}}(t_{i})-S_{a^{-}_{0}}(t_{i})],\\
&E_{af}(t_{i})=\alpha_{s}(t_{i})\exp[-S_{a^{+}_{0}}(t_{i})-S_{a^{-}_{0}}(t_{i})],\\
&E_{nef}(t_{i})=\alpha_{s}(t_{i})\exp[-S_{B_{s}}(t_{i})-S_{a^{+}_{0}}(t_{i})-S_{a^{-}_{0}}(t_{i})]_{\emph{b}_{1}=\emph{b}_{3}},\\
&E_{naf}(t_{i})=\alpha_{s}(t_{i})\exp[-S_{B_{s}}(t_{i})-S_{a^{+}_{0}}(t_{i})-S_{a^{-}_{0}}(t_{i})]_{\emph{b}_{2}=\emph{b}_{3}}.
\end{split}
\end{equation}
where the largest energy scales $t_{i} (i=a,c,e,g)$ to eliminate the large logarithmic radiative corrections are chosen as:
\begin{equation}
\begin{split}
&t^1_{a}=\mathrm{max}\{M_{B_{s}}\sqrt{1-\emph{x}_{2}}, 1/\emph{b}_{1}, 1/\emph{b}_{2}\},\\
&t^2_{a}=\mathrm{max}\{M_{B_{s}}\sqrt{\emph{x}_{1}}, 1/\emph{b}_{1}, 1/\emph{b}_{2}\},\\
&t^1_{c}=\mathrm{max}\{M_{B_{s}}\sqrt{|\emph{x}_{1}+\emph{x}_{2}+\emph{x}_{3}-\emph{x}_{1}\emph{x}_{2}-\emph{x}_{2}\emph{x}_{3}-1|}, M_{B_{s}}\sqrt{\emph{x}_{1}(1-\emph{x}_{2})}, 1/\emph{b}_{2}, 1/\emph{b}_{3}\},\\
&t^2_{c}=\mathrm{max}\{M_{B_{s}}\sqrt{|\emph{x}_{1}-\emph{x}_{3}-\emph{x}_{1}\emph{x}_{2}+\emph{x}_{2}\emph{x}_{3}|}, M_{B_{s}}\sqrt{\emph{x}_{1}(1-\emph{x}_{2})}, 1/\emph{b}_{2}, 1/\emph{b}_{3}\},\\
&t^1_{e}=\mathrm{max}\{M_{B_{s}}\sqrt{\emph{x}_{3}}, 1/\emph{b}_{2}, 1/\emph{b}_{3}\},\\
&t^2_{e}=\mathrm{max}\{M_{B_{s}}\sqrt{\emph{x}_{2}}, 1/\emph{b}_{2}, 1/\emph{b}_{3}\},\\
&t^{1}_{g}=\mathrm{max}\{M_{B_{s}}\sqrt{\emph{x}_{2}\emph{x}_{3}}, M_{B_{s}}\sqrt{|\emph{x}_{1}\emph{x}_{2}-\emph{x}_{2}\emph{x}_{3}|}, 1/\emph{b}_{1}, 1/\emph{b}_{2}\},\\
&t^{2}_{g}=\mathrm{max}\{M_{B_{s}}\sqrt{\emph{x}_{2}\emph{x}_{3}}, M_{B_{s}}\sqrt{|\emph{x}_{1}+\emph{x}_{2}+\emph{x}_{3}-\emph{x}_{1}\emph{x}_{2}-\emph{x}_{2}\emph{x}_{3}|}, 1/\emph{b}_{1}, 1/\emph{b}_{2}\}.
\end{split}
\end{equation}

The $S_{B_{s}}(\emph{x}_{1})$, $S_{S}(\emph{x}_{i})$ used in the decay amplitudes are defined as:
\begin{eqnarray}
\begin{split}
&S_{B_{s}}(\emph{x}_{1})=s(\emph{x}_{1}p^+_{1},\emph{b}_{1})+\frac{5}{3}\int^{t}_{1/\emph{b}_{1}}\frac{\emph{d}{\bar{\mu}}}{\bar{\mu}}\gamma_{q}(\alpha_{s}({\bar{\mu}})),\\
&S_{S}(\emph{x}_{2})=s(\emph{x}_{2}p^{+}_{2},\emph{b}_{2})+s(\bar{\emph{x}}_{2}p^{+}_{2},\emph{b}_{2})+2\int^{t}_{1/\emph{b}_{2}}\frac{\emph{d}{\bar{\mu}}}{\bar{\mu}}\gamma_{q}(\alpha_{s}({\bar{\mu}})),\\
&S_{S}(\emph{x}_{3})=s(\emph{x}_{3}p^{-}_{3},\emph{b}_{3})+s(\bar{\emph{x}}_{3}p^-_{3},\emph{b}_{3})+2\int^{t}_{1/\emph{b}_{3}}\frac{\emph{d}{\bar{\mu}}}{\bar{\mu}}\gamma_{q}(\alpha_{s}({\bar{\mu}})),
\end{split}
\end{eqnarray}
where $\bar{\emph{x}}_{i}=1-\emph{x}_{i}$ and $\gamma_{q}=-\alpha_{s}/\pi$ is the anomalous dimension of the quark, and the Sudakov factor $s(Q,\emph{b})$ are resulting from the resummation of double logarithms and can be found in Ref.~\cite{Li:2002mi},

\begin{equation}
s(Q,\emph{b})=\int^{Q}_{1/\emph{b}}\frac{d\mu}{\mu}[\ln(\frac{Q}{\mu})A(\alpha\bar{(\mu)})+B(\alpha_{s}\bar{(\mu)})]
\end{equation}
with
\begin{equation}
\begin{split}
&A=C_{F}\frac{\alpha_{s}}{\pi}+[\frac{67}{9}-\frac{\pi^2}{3}-\frac{10}{27}n_{f}+\frac{3}{2}\beta_{0}\ln(\frac{e^{\gamma_{E}}}{2})](\frac{\alpha_{s}}{\pi})^2,\\
\\
&B=\frac{2}{3}\frac{\alpha_{s}}{\pi}\ln(\frac{e^{2\gamma_{E}-1}}{2}),
\end{split}
\end{equation}
where $\gamma_{E}$ and $n_{f}$ are Euler constant and the active flavor number, respectively.

The threshold resummation factor $S_{t}(\emph{x})$ have been parameterized in~\cite{Kurimoto:2001zj}, which is:
\begin{equation}
S_{t}(\emph{x})=\frac{2^{1+2c}\Gamma(\frac{3}{2}+c)}{\sqrt{\pi}\Gamma(1+c)}[\emph{x}(1-\emph{x})]^{c}
\end{equation}
with the fitted parameter $c=0.3$.

%%%--=================================================================
%%%=====           Reference        ==========================
%%5===================================================================

\end{document}